\magnification=\magstep1
\vsize=47pc


\catcode`\@=11


\message{Loading a modification of the jyTeX macros...}

\message{modifications to plain.tex,}


\def\newcount{\alloc@0\count\countdef\insc@unt}
\def\newdimen{\alloc@1\dimen\dimendef\insc@unt}
\def\newskip{\alloc@2\skip\skipdef\insc@unt}
\def\newmuskip{\alloc@3\muskip\muskipdef\@cclvi}
\def\newtoks{\alloc@5\toks\toksdef\@cclvi}
\def\newhelp#1#2{\newtoks#1\global#1\expandafter{\csname#2\endcsname}}
\def\newread{\alloc@6\read\chardef\sixt@@n}
\def\newwrite{\alloc@7\write\chardef\sixt@@n}
\def\newfam{\alloc@8\fam\chardef\sixt@@n}
\def\newinsert#1{\global\advance\insc@unt by\m@ne
     \ch@ck0\insc@unt\count
     \ch@ck1\insc@unt\dimen
     \ch@ck2\insc@unt\skip
     \ch@ck4\insc@unt\box
     \allocationnumber=\insc@unt
     \global\chardef#1=\allocationnumber
     \wlog{\string#1=\string\insert\the\allocationnumber}}
\def\newif#1{\count@\escapechar \escapechar\m@ne
     \expandafter\expandafter\expandafter
          \xdef\@if#1{true}{\let\noexpand#1=\noexpand\iftrue}%
     \expandafter\expandafter\expandafter
          \xdef\@if#1{false}{\let\noexpand#1=\noexpand\iffalse}%
     \global\@if#1{false}\escapechar=\count@}


\newlinechar=`\^^J
\overfullrule=0pt

\message{hacks,}


\toksdef\toks@i=1
\toksdef\toks@ii=2


\def\TeX{T\kern-.1667em \lower.5ex \hbox{E}\kern-.125em X\null}
\def\jyTeX{{\leavevmode
     \raise.587ex \hbox{\it\j}\kern-.1em \lower.048ex \hbox{\it y}\kern-.12em
     \TeX}}

\let\then=\iftrue
\def\ifnoarg#1\then{\def\hack@{#1}\ifx\hack@\empty}
\def\ifundefined#1\then{%
     \expandafter\ifx\csname\expandafter\blank\string#1\endcsname\relax}
\def\useif#1\then{\csname#1\endcsname}
\def\usename#1{\csname#1\endcsname}
\def\useafter#1#2{\expandafter#1\csname#2\endcsname}

\long\def\loop#1\repeat{\def\@iterate{#1\expandafter\@iterate\fi}\@iterate
     \let\@iterate=\relax}

\let\TeXend=\end
\def\begin#1{\begingroup\def\@@blockname{#1}\usename{begin#1}}
\def\End#1{\usename{end#1}\def\hack@{#1}%
     \ifx\@@blockname\hack@
          \endgroup
     \else\err@badgroup\hack@\@@blockname
     \fi}
\def\@@blockname{}

\def\defaultoption[#1]#2{%
     \def\hack@{\ifx\hack@ii[\toks@={#2}\else\toks@={#2[#1]}\fi\the\toks@}%
     \futurelet\hack@ii\hack@}

\def\markup#1{\let\@@marksf=\empty
     \ifhmode\edef\@@marksf{\spacefactor=\the\spacefactor\relax}\/\fi
     ${}^{\hbox{\subscriptfonts#1}}$\@@marksf}


\newtoks\shortyear
\newtoks\militaryhour
\newtoks\standardhour
\newtoks\minute
\newtoks\amorpm

\def\settime{\count@=\time\divide\count@ by60
     \militaryhour=\expandafter{\number\count@}%
     {\multiply\count@ by-60 \advance\count@ by\time
          \xdef\hack@{\ifnum\count@<10 0\fi\number\count@}}%
     \minute=\expandafter{\hack@}%
     \ifnum\count@<12
          \amorpm={am}
     \else\amorpm={pm}
          \ifnum\count@>12 \advance\count@ by-12 \fi
     \fi
     \standardhour=\expandafter{\number\count@}%
     \def\hack@19##1##2{\shortyear={##1##2}}%
          \expandafter\hack@\the\year}

\def\monthword#1{%
     \ifcase#1
          $\bullet$\err@badcountervalue{monthword}%
          \or January\or February\or March\or April\or May\or June%
          \or July\or August\or September\or October\or November\or December%
     \else$\bullet$\err@badcountervalue{monthword}%
     \fi}

\def\monthabbr#1{%
     \ifcase#1
          $\bullet$\err@badcountervalue{monthabbr}%
          \or Jan\or Feb\or Mar\or Apr\or May\or Jun%
          \or Jul\or Aug\or Sep\or Oct\or Nov\or Dec%
     \else$\bullet$\err@badcountervalue{monthabbr}%
     \fi}

\def\militarytime{\the\militaryhour:\the\minute}
\def\standardtime{\the\standardhour:\the\minute}


\def\@setnumstyle#1#2{\expandafter\global\expandafter\expandafter
     \expandafter\let\expandafter\expandafter
     \csname @\expandafter\blank\string#1style\endcsname
     \csname#2\endcsname}
\def\numstyle#1{\usename{@\expandafter\blank\string#1style}#1}
\def\ifblank#1\then{\useafter\ifx{@\expandafter\blank\string#1}\blank}

\def\blank#1{}

\def\Roman#1{\expandafter\uppercase\expandafter{\romannumeral#1}}
\def\alphabetic#1{%
     \ifcase#1
          $\bullet$\err@badcountervalue{alphabetic}%
          \or a\or b\or c\or d\or e\or f\or g\or h\or i\or j\or k\or l\or m%
          \or n\or o\or p\or q\or r\or s\or t\or u\or v\or w\or x\or y\or z%
     \else$\bullet$\err@badcountervalue{alphabetic}%
     \fi}
\def\Alphabetic#1{\expandafter\uppercase\expandafter{\alphabetic{#1}}}
\def\symbols#1{%
     \ifcase#1
          $\bullet$\err@badcountervalue{symbols}%
          \or*\or\dag\or\ddag\or\S\or$\|$%
          \or**\or\dag\dag\or\ddag\ddag\or\S\S\or$\|\|$%
     \else$\bullet$\err@badcountervalue{symbols}%
     \fi}


\catcode`\^^?=13 \def^^?{\relax}

\def\trimleading#1\to#2{\edef#2{#1}%
     \expandafter\@trimleading\expandafter#2#2^^?^^?}
\def\@trimleading#1#2#3^^?{\ifx#2^^?\def#1{}\else\def#1{#2#3}\fi}

\def\trimtrailing#1\to#2{\edef#2{#1}%
     \expandafter\@trimtrailing\expandafter#2#2^^? ^^?\relax}
\def\@trimtrailing#1#2 ^^?#3{\ifx#3\relax\toks@={}%
     \else\def#1{#2}\toks@={\trimtrailing#1\to#1}\fi
     \the\toks@}

\def\trim#1\to#2{\trimleading#1\to#2\trimtrailing#2\to#2}

\catcode`\^^?=15


\long\def\additemL#1\to#2{\toks@={\^^\{#1}}\toks@ii=\expandafter{#2}%
     \xdef#2{\the\toks@\the\toks@ii}}

\long\def\additemR#1\to#2{\toks@={\^^\{#1}}\toks@ii=\expandafter{#2}%
     \xdef#2{\the\toks@ii\the\toks@}}

\def\getitemL#1\to#2{\expandafter\@getitemL#1\hack@#1#2}
\def\@getitemL\^^\#1#2\hack@#3#4{\def#4{#1}\def#3{#2}}


\newskip\headskip
\newskip\footskip

\message{document layout,}

\newif\ifdraft
\def\draft{\drafttrue\leftmargin=.5in \overfullrule=5pt }


\newskip\abovechapterskip
\newskip\belowchapterskip
\newskip\abovesectionskip
\newskip\belowsectionskip
\newskip\abovesubsectionskip
\newskip\belowsubsectionskip

\def\chapterstyle#1{\global\expandafter\let\expandafter\@chapterstyle
     \csname#1text\endcsname}
\def\sectionstyle#1{\global\expandafter\let\expandafter\@sectionstyle
     \csname#1text\endcsname}
\def\subsectionstyle#1{\global\expandafter\let\expandafter\@subsectionstyle
     \csname#1text\endcsname}

\def\CHapter#1{%
     \ifdim\lastskip=17sp \else\chapterbreak\vskip\abovechapterskip\fi
     \@chapterstyle{\ifblank\chapternumstyle\then
          \else\newchapternum=\next\chapternumformat\ \fi#1}%
     \nobreak\vskip\belowchapterskip\vskip17sp }

\def\Section#1{%
     \ifdim\lastskip=17sp \else\sectionbreak\vskip\abovesectionskip\fi
     \@sectionstyle{\ifblank\sectionnumstyle\then
          \else\newsectionnum=\next\sectionnumformat\ \fi#1}%
     \nobreak\vskip\belowsectionskip\vskip17sp }

\def\subsection#1{%
     \ifdim\lastskip=17sp \else\subsectionbreak\vskip\abovesubsectionskip\fi
     \@subsectionstyle{\ifblank\subsectionnumstyle\then
          \else\newsubsectionnum=\next\subsectionnumformat\ \fi#1}%
     \nobreak\vskip\belowsubsectionskip\vskip17sp }


\newtoks\everybye \everybye={\par\vfil}
\outer\def\bye{\the\everybye
     \footnotecheck
     \prelabelcheck
     \streamcheck
     \supereject
     \TeXend}

\message{labels,}

\let\@@labeldef=\xdef
\newif\if@labelfile
\newwrite\@labelfile
\let\@prelabellist=\empty

\def\Label#1#2{\trim#1\to\@@labarg\edef\@@labtext{#2}%
     \edef\@@labname{lab@\@@labarg}%
     \useafter\ifundefined\@@labname\then\else\@yeslab\fi
     \useafter\@@labeldef\@@labname{#2}%
     \ifstreaming
          \expandafter\toks@\expandafter\expandafter\expandafter
               {\csname\@@labname\endcsname}%
          \immediate\write\streamout{\noexpand\Label{\@@labarg}{\the\toks@}}%
     \fi}
\def\@yeslab{%
     \useafter\ifundefined{if\@@labname}\then
          \err@labelredef\@@labarg
     \else\useif{if\@@labname}\then
               \err@labelredef\@@labarg
          \else\global\usename{\@@labname true}%
               \useafter\ifundefined{pre\@@labname}\then
               \else\useafter\ifx{pre\@@labname}\@@labtext
                    \else\err@badlabelmatch\@@labarg
                    \fi
               \fi
               \if@labelfile
               \else\global\@labelfiletrue
                    \immediate\write\sixt@@n{--> Creating file \jobname.lab}%
                    \immediate\openout\@labelfile=\jobname.lab
               \fi
               \immediate\write\@labelfile
                    {\noexpand\prelabel{\@@labarg}{\@@labtext}}%
          \fi
     \fi}

\def\putlab#1{\trim#1\to\@@labarg\edef\@@labname{lab@\@@labarg}%
     \useafter\ifundefined\@@labname\then\@nolab\else\usename\@@labname\fi}
\def\@nolab{%
     \useafter\ifundefined{pre\@@labname}\then
          \undefinedlabelformat
          \err@needlabel\@@labarg
          \useafter\xdef\@@labname{\undefinedlabelformat}%
     \else\usename{pre\@@labname}%
          \useafter\xdef\@@labname{\usename{pre\@@labname}}%
     \fi
     \useafter\newif{if\@@labname}%
     \expandafter\additemR\@@labarg\to\@prelabellist}

\def\prelabel#1{\useafter\gdef{prelab@#1}}

\def\ifundefinedlabel#1\then{%
     \expandafter\ifx\csname lab@#1\endcsname\relax}
\def\useiflab#1\then{\csname iflab@#1\endcsname}

\def\prelabelcheck{{%
     \def\^^\##1{\useiflab{##1}\then\else\err@undefinedlabel{##1}\fi}%
     \@prelabellist}}

\message{equation numbering,}

\newcount\chapternum
\newcount\sectionnum
\newcount\subsectionnum
\newcount\equationnum
\newcount\subequationnum
\newcount\figurenum
\newcount\subfigurenum
\newcount\tablenum
\newcount\subtablenum
\newcount\defnum
\newcount\subdefnum
\newcount\thmnum
\newcount\subthmnum
\newcount\lemnum
\newcount\sublemnum

\newif\if@subeqncount
\newif\if@subfigcount
\newif\if@subtblcount
\newif\if@subdefcount
\newif\if@subthmcount
\newif\if@sublemcount

\def\newchapternum{\newsectionnum=\z@\@resetnum\chapternum}
\def\newsectionnum{\newsubsectionnum=\z@\@resetnum\sectionnum}
\def\newsubsectionnum{\newequationnum=\z@\newfigurenum=\z@\newtablenum=\z@
     \newdefnum=\z@\newthmnum=\z@\newlemnum=\z@
     \@resetnum\subsectionnum}
\def\newequationnum{\newsubequationnum=\z@\@resetnum\equationnum}
\def\newsubequationnum{\@resetnum\subequationnum}
\def\newfigurenum{\newsubfigurenum=\z@\@resetnum\figurenum}
\def\newsubfigurenum{\@resetnum\subfigurenum}
\def\newtablenum{\newsubtablenum=\z@\@resetnum\tablenum}
\def\newsubtablenum{\@resetnum\subtablenum}
\def\newdefnum{\newsubdefnum=\z@\@resetnum\defnum}
\def\newsubdefnum{\@resetnum\subdefnum}
\def\newthmnum{\newsubthmnum=\z@\@resetnum\thmnum}
\def\newsubthmnum{\@resetnum\subthmnum}
\def\newlemnum{\newsublemnum=\z@\@resetnum\lemnum}
\def\newsublemnum{\@resetnum\sublemnum}

\def\@resetnum#1{\global\advance#1by1 \edef\next{\the#1\relax}\global#1}

\newchapternum=0

\def\chapternumstyle#1{\@setnumstyle\chapternum{#1}}
\def\sectionnumstyle#1{\@setnumstyle\sectionnum{#1}}
\def\subsectionnumstyle#1{\@setnumstyle\subsectionnum{#1}}
\def\equationnumstyle#1{\@setnumstyle\equationnum{#1}}
\def\subequationnumstyle#1{\@setnumstyle\subequationnum{#1}%
     \ifblank\subequationnumstyle\then\global\@subeqncountfalse\fi
     \ignorespaces}
\def\figurenumstyle#1{\@setnumstyle\figurenum{#1}}
\def\subfigurenumstyle#1{\@setnumstyle\subfigurenum{#1}%
     \ifblank\subfigurenumstyle\then\global\@subfigcountfalse\fi
     \ignorespaces}
\def\tablenumstyle#1{\@setnumstyle\tablenum{#1}}
\def\subtablenumstyle#1{\@setnumstyle\subtablenum{#1}%
     \ifblank\subtablenumstyle\then\global\@subtblcountfalse\fi
     \ignorespaces}
\def\defnumstyle#1{\@setnumstyle\defnum{#1}}
\def\subdefnumstyle#1{\@setnumstyle\subdefnum{#1}%
     \ifblank\subdefnumstyle\then\global\@subdefcountfalse\fi
     \ignorespaces}
\def\thmnumstyle#1{\@setnumstyle\thmnum{#1}}
\def\subthmnumstyle#1{\@setnumstyle\subthmnum{#1}%
     \ifblank\subthmnumstyle\then\global\@subthmcountfalse\fi
     \ignorespaces}
\def\lemnumstyle#1{\@setnumstyle\lemnum{#1}}
\def\sublemnumstyle#1{\@setnumstyle\sublemnum{#1}%
     \ifblank\sublemnumstyle\then\global\@sublemcountfalse\fi
     \ignorespaces}

\def\heqnlabel{\newequationnum=\next
          \ifblank\subequationnumstyle\then
          \else\global\@subeqncounttrue
               \newsubequationnum=\@ne
          \fi}

\def\eqnlabel#1{%
     \if@subeqncount
          \newsubequationnum=\next
     \else\heqnlabel
     \fi
     \Label{#1}{\puteqnformat}(\puteqn{#1})%
     \ifdraft\rlap{\hskip.1in{\tt#1}}\fi}

\let\puteqn=\putlab

\def\putequation#1{\useafter\ifundefined{eqn@#1}\then
     \err@undefinedeqn{#1}\else\usename{eqn@#1}\fi}

\def\eqnseriesstyle#1{\gdef\@eqnseriesstyle{#1}}
\def\begineqnseries{\subequationnumstyle{\@eqnseriesstyle}%
     \defaultoption[]\@begineqnseries}
\def\@begineqnseries[#1]{\edef\@@eqnname{#1}}
\def\endeqnseries{\subequationnumstyle{blank}%
     \expandafter\ifnoarg\@@eqnname\then
     \else\Label\@@eqnname{\puteqnformat}%
     \fi
     \aftergroup\ignorespaces}

\def\figlabel#1{%
     \if@subfigcount
          \newsubfigurenum=\next
     \else\newfigurenum=\next
          \ifblank\subfigurenumstyle\then
          \else\global\@subfigcounttrue
               \newsubfigurenum=\@ne
          \fi
     \fi
     \Label{#1}{\putfigformat}\putfig{#1}%
   }

\let\putfig=\putlab

\def\figseriesstyle#1{\gdef\@figseriesstyle{#1}}
\def\beginfigseries{\subfigurenumstyle{\@figseriesstyle}%
     \defaultoption[]\@beginfigseries}
\def\@beginfigseries[#1]{\edef\@@figname{#1}}
\def\endfigseries{\subfigurenumstyle{blank}%
     \expandafter\ifnoarg\@@figname\then
     \else\Label\@@figname{\putfigformat}%
     \fi
     \aftergroup\ignorespaces}

\def\tbllabel#1{%
     \if@subtblcount
          \newsubtablenum=\next
     \else\newtablenum=\next
          \ifblank\subtablenumstyle\then
          \else\global\@subtblcounttrue
               \newsubtablenum=\@ne
          \fi
     \fi
     \Label{#1}{\puttblformat}\puttbl{#1}%
}

\let\puttbl=\putlab

\def\tblseriesstyle#1{\gdef\@tblseriesstyle{#1}}
\def\begintblseries{\subtablenumstyle{\@tblseriesstyle}%
     \defaultoption[]\@begintblseries}
\def\@begintblseries[#1]{\edef\@@tblname{#1}}
\def\endtblseries{\subtablenumstyle{blank}%
     \expandafter\ifnoarg\@@tblname\then
     \else\Label\@@tblname{\puttblformat}%
     \fi
     \aftergroup\ignorespaces}


\def\deflab#1{%
     \if@subdefcount
          \newsubdefnum=\next
     \else\newdefnum=\next
          \ifblank\subdefnumstyle\then
          \else\global\@subdefcounttrue
               \newsubdefnum=\@ne
          \fi
     \fi
     \Label{#1}{\putdefformat}\refdef{#1}%
}

\let\refdef=\putlab

\def\defseriesstyle#1{\gdef\@defseriesstyle{#1}}
\def\begindefseries{\subtablenumstyle{\@defseriesstyle}%
     \defaultoption[]\@begindefseries}
\def\@begindefseries[#1]{\edef\@@defname{#1}}
\def\enddefseries{\subdefnumstyle{blank}%
     \expandafter\ifnoarg\@@defname\then
     \else\Label\@@defname{\putdefformat}%
     \fi
     \aftergroup\ignorespaces}

\def\thmlab#1{%
     \if@subthmcount
          \newsubthmnum=\next
     \else\newthmnum=\next
          \ifblank\subthmnumstyle\then
          \else\global\@subthmcounttrue
               \newsubthmnum=\@ne
          \fi
     \fi
     \Label{#1}{\putthmformat}\refthm{#1}%
}

\let\refthm=\putlab

\def\thmseriesstyle#1{\gdef\@thmseriesstyle{#1}}
\def\beginthmseries{\subthmnumstyle{\@thmseriesstyle}%
     \defaultoption[]\@beginthmseries}
\def\@beginthmseries[#1]{\edef\@@thmname{#1}}
\def\endthmseries{\subthmstyle{blank}%
     \expandafter\ifnoarg\@@thmname\then
     \else\Label\@@thmname{\putthmformat}%
     \fi
     \aftergroup\ignorespaces}

\def\lemlab#1{%
     \if@sublemcount
          \newsublemnum=\next
     \else\newlemnum=\next
          \ifblank\sublemnumstyle\then
          \else\global\@sublemcounttrue
               \newsublemnum=\@ne
          \fi
     \fi
     \Label{#1}{\putlemformat}\reflem{#1}%
}

\let\reflem=\putlab

\def\lemseriesstyle#1{\gdef\@lemseriesstyle{#1}}
\def\beginlemseries{\sublemnumstyle{\@lemseriesstyle}%
     \defaultoption[]\@beginlemseries}
\def\@beginlemseries[#1]{\edef\@@lemname{#1}}
\def\endlemseries{\sublemnumstyle{blank}%
     \expandafter\ifnoarg\@@lemname\then
     \else\Label\@@lemname{\putlemformat}%
     \fi
     \aftergroup\ignorespaces}

\message{reference numbering,}

\newcount\referencenum \referencenum=0
\newcount\@@prerefcount \@@prerefcount=0
\newcount\@@thisref
\newcount\@@lastref
\newcount\@@loopref
\newcount\@@refseq
\newdimen\refnumindent
\let\@undefreflist=\empty

\def\referencenumstyle#1{\@setnumstyle\referencenum{#1}}

\def\referencestyle#1{\usename{@ref#1}}

\def\@refsequential{%
     \gdef\@refpredef##1{\global\advance\referencenum by\@ne
          \let\^^\=0\Label{##1}{\^^\{\the\referencenum}}%
          \useafter\gdef{ref@\the\referencenum}{{##1}{\undefinedlabelformat}}}%
     \gdef\@reference##1##2{%
          \ifundefinedlabel##1\then
          \else\def\^^\####1{\global\@@thisref=####1\relax}\putlab{##1}%
               \useafter\gdef{ref@\the\@@thisref}{{##1}{##2}}%
          \fi}%
     \gdef\endputreferences{%
          \loop\ifnum\@@loopref<\referencenum
                    \advance\@@loopref by\@ne
                    \expandafter\expandafter\expandafter\@printreference
                         \csname ref@\the\@@loopref\endcsname
          \repeat
          \par}}

\def\@refpreordered{%
     \gdef\@refpredef##1{\global\advance\referencenum by\@ne
          \additemR##1\to\@undefreflist}%
     \gdef\@reference##1##2{%
          \ifundefinedlabel##1\then
          \else\global\advance\@@loopref by\@ne
               {\let\^^\=0\Label{##1}{\^^\{\the\@@loopref}}}%
               \@printreference{##1}{##2}%
          \fi}
     \gdef\endputreferences{%
          \def\^^\####1{\useiflab{####1}\then
               \else\reference{####1}{\undefinedlabelformat}\fi}%
          \@undefreflist
          \par}}

\def\beginprereferences{\par
     \def\reference##1##2{\global\advance\referencenum by1\@ne
          \let\^^\=0\Label{##1}{\^^\{\the\referencenum}}%
          \useafter\gdef{ref@\the\referencenum}{{##1}{##2}}}}
\def\endprereferences{\global\@@prerefcount=\the\referencenum\par}

\def\beginputreferences{\par
     \refnumindent=\z@\@@loopref=\z@
     \loop\ifnum\@@loopref<\referencenum
               \advance\@@loopref by\@ne
               \setbox\z@=\hbox{\referencenum=\@@loopref
                    \referencenumformat\enskip}%
               \ifdim\wd\z@>\refnumindent\refnumindent=\wd\z@\fi
     \repeat
     \putreferenceformat
     \@@loopref=\z@
     \loop\ifnum\@@loopref<\@@prerefcount
               \advance\@@loopref by\@ne
               \expandafter\expandafter\expandafter\@printreference
                    \csname ref@\the\@@loopref\endcsname
     \repeat
     \let\reference=\@reference}

\def\@printreference#1#2{\ifx#2\undefinedlabelformat\err@undefinedref{#1}\fi
     \noindent\ifdraft\rlap{\hskip\hsize\hskip.1in \tt#1}\fi
     \llap{\referencenum=\@@loopref\referencenumformat\enskip}#2\par}

\def\reference#1#2{{\par\refnumindent=\z@\putreferenceformat\noindent#2\par}}

\def\putref#1{\trim#1\to\@@refarg
     \expandafter\ifnoarg\@@refarg\then
          \toks@={\relax}%
     \else\@@lastref=-\@m\def\@@refsep{}\def\@more{\@nextref}%
          \toks@={\@nextref#1,,}%
     \fi\the\toks@}
\def\@nextref#1,{\trim#1\to\@@refarg
     \expandafter\ifnoarg\@@refarg\then
          \let\@more=\relax
     \else\ifundefinedlabel\@@refarg\then
               \expandafter\@refpredef\expandafter{\@@refarg}%
          \fi
          \def\^^\##1{\global\@@thisref=##1\relax}%
          \global\@@thisref=\m@ne
          \setbox\z@=\hbox{\putlab\@@refarg}%
     \fi
     \advance\@@lastref by\@ne
     \ifnum\@@lastref=\@@thisref\advance\@@refseq by\@ne\else\@@refseq=\@ne\fi
     \ifnum\@@lastref<\z@
     \else\ifnum\@@refseq<\thr@@
               \@@refsep\def\@@refsep{,}%
               \ifnum\@@lastref>\z@
                    \advance\@@lastref by\m@ne
                    {\referencenum=\@@lastref\putrefformat}%
               \else\undefinedlabelformat
               \fi
          \else\def\@@refsep{--}%
          \fi
     \fi
     \@@lastref=\@@thisref
     \@more}

\message{streaming,}

\newif\ifstreaming

\def\streamto{\defaultoption[\jobname]\@streamto}
\def\@streamto[#1]{\global\streamingtrue
     \immediate\write\sixt@@n{--> Streaming to #1.str}%
     \newwrite\streamout\immediate\openout\streamout=#1.str }

\def\streamfrom{\defaultoption[\jobname]\@streamfrom}
\def\@streamfrom[#1]{\newread\streamin\openin\streamin=#1.str
     \ifeof\streamin
          \expandafter\err@nostream\expandafter{#1.str}%
     \else\immediate\write\sixt@@n{--> Streaming from #1.str}%
          \let\@@labeldef=\gdef
          \ifstreaming
               \edef\@elc{\endlinechar=\the\endlinechar}%
               \endlinechar=\m@ne
               \loop\read\streamin to\@@scratcha
                    \ifeof\streamin
                         \streamingfalse
                    \else\toks@=\expandafter{\@@scratcha}%
                         \immediate\write\streamout{\the\toks@}%
                    \fi
                    \ifstreaming
               \repeat
               \@elc
               \input #1.str
               \streamingtrue
          \else\input #1.str
          \fi
          \let\@@labeldef=\xdef
     \fi}

\def\streamcheck{\ifstreaming
     \immediate\write\streamout{\pagenum=\the\pagenum}%
     \immediate\write\streamout{\footnotenum=\the\footnotenum}%
     \immediate\write\streamout{\referencenum=\the\referencenum}%
     \immediate\write\streamout{\chapternum=\the\chapternum}%
     \immediate\write\streamout{\sectionnum=\the\sectionnum}%
     \immediate\write\streamout{\subsectionnum=\the\subsectionnum}%
     \immediate\write\streamout{\equationnum=\the\equationnum}%
     \immediate\write\streamout{\subequationnum=\the\subequationnum}%
     \immediate\write\streamout{\figurenum=\the\figurenum}%
     \immediate\write\streamout{\subfigurenum=\the\subfigurenum}%
     \immediate\write\streamout{\tablenum=\the\tablenum}%
     \immediate\write\streamout{\subtablenum=\the\subtablenum}%
     \immediate\closeout\streamout
     \fi}


\def\err@badtypesize{%
     \errhelp={The limited availability of certain fonts requires^^J%
          that the base type size be 10pt, 12pt, or 14pt.^^J}%
     \errmessage{--> Illegal base type size}}

\def\err@badsizechange{\immediate\write\sixt@@n
     {--> Size change not allowed in math mode, ignored}}

\def\err@sizetoolarge#1{\immediate\write\sixt@@n
     {--> \noexpand#1 too big, substituting HUGE}}

\def\err@sizenotavailable#1{\immediate\write\sixt@@n
     {--> Size not available, \noexpand#1 ignored}}

\def\err@fontnotavailable#1{\immediate\write\sixt@@n
     {--> Font not available, \noexpand#1 ignored}}

\def\err@sltoit{\immediate\write\sixt@@n
     {--> Style \noexpand\sl not available, substituting \noexpand\it}%
     \it}

\def\err@bfstobf{\immediate\write\sixt@@n
     {--> Style \noexpand\bfs not available, substituting \noexpand\bf}%
     \bf}

\def\err@badgroup#1#2{%
     \errhelp={The block you have just tried to close was not the one^^J%
          most recently opened.^^J}%
     \errmessage{--> \noexpand\End{#1} doesn't match \noexpand\begin{#2}}}

\def\err@badcountervalue#1{\immediate\write\sixt@@n
     {--> Counter (#1) out of bounds}}

\def\err@extrafootnotemark{\immediate\write\sixt@@n
     {--> \noexpand\footnotemark command
          has no corresponding \noexpand\footnotetext}}

\def\err@extrafootnotetext{%
     \errhelp{You have given a \noexpand\footnotetext command without first
          specifying^^Ja \noexpand\footnotemark.^^J}%
     \errmessage{--> \noexpand\footnotetext command has no corresponding
          \noexpand\footnotemark}}

\def\err@labelredef#1{\immediate\write\sixt@@n
     {--> Label "#1" redefined}}

\def\err@badlabelmatch#1{\immediate\write\sixt@@n
     {--> Definition of label "#1" doesn't match value in \jobname.lab}}

\def\err@needlabel#1{\immediate\write\sixt@@n
     {--> Label "#1" cited before its definition}}

\def\err@undefinedlabel#1{\immediate\write\sixt@@n
     {--> Label "#1" cited but never defined}}

\def\err@undefinedeqn#1{\immediate\write\sixt@@n
     {--> Equation "#1" not defined}}

\def\err@undefinedref#1{\immediate\write\sixt@@n
     {--> Reference "#1" not defined}}

\def\err@nostream#1{%
     \errhelp={You have tried to input a stream file that doesn't exist.^^J}%
     \errmessage{--> Stream file #1 not found}}

\message{jyTeX initialization}

\everyjob{\immediate\write16{--> jyTeX version \fmtversion}%
     \edef\@@jobname{\jobname}%
     \edef\jobname{\@@jobname}%
     \settime
     \openin0=\jobname.lab
     \ifeof0
     \else\closein0
          \immediate\write16{--> Getting labels from file \jobname.lab}%
          \input\jobname.lab
     \fi}


%
     \^^\{\splittopskip}%
     \^^\{\maxdepth}%
     \^^\{\skip\topins}%
     \^^\{\skip\footins}%
     \^^\{\headskip}%
     \^^\{\footskip}}

\def\scalingskipslist{%
     \^^\{\p@renwd}%
     \^^\{\delimitershortfall}%
     \^^\{\nulldelimiterspace}%
     \^^\{\scriptspace}%
     \^^\{\jot}%
     \^^\{\normalbaselineskip}%
     \^^\{\normallineskip}%
     \^^\{\normallineskiplimit}%
     \^^\{\baselineskip}%
     \^^\{\lineskip}%
     \^^\{\lineskiplimit}%
     \^^\{\bigskipamount}%
     \^^\{\medskipamount}%
     \^^\{\smallskipamount}%
     \^^\{\parskip}%
     \^^\{\parindent}%
     \^^\{\abovedisplayskip}%
     \^^\{\belowdisplayskip}%
     \^^\{\abovedisplayshortskip}%
     \^^\{\belowdisplayshortskip}%
     \^^\{\abovechapterskip}%
     \^^\{\belowchapterskip}%
     \^^\{\abovesectionskip}%
     \^^\{\belowsectionskip}%
     \^^\{\abovesubsectionskip}%
     \^^\{\belowsubsectionskip}}


\def\twoupsetup{
     \topmargin=.75in
     \leftmargin=.5in
     \vsize=6.9in
     \hsize=4.75in
     \fullhsize=10in
     \let\draft=\relax}


\chapterstyle{left}                              
\chapternumstyle{blank}                          
\def\chapterbreak{\newpage}                      
\abovechapterskip=0pt                            
\belowchapterskip=1.5\baselineskip               
     plus.38\baselineskip minus.38\baselineskip
\def\chapternumformat{\numstyle\chapternum.}     

\sectionstyle{left}                              
\sectionnumstyle{blank}                          
\def\sectionbreak{\vskip0pt plus4\baselineskip\penalty-100
     \vskip0pt plus-4\baselineskip}              
\abovesectionskip=1.5\baselineskip               
     plus.38\baselineskip minus.38\baselineskip
\belowsectionskip=\the\baselineskip              
     plus.25\baselineskip minus.25\baselineskip
\def\sectionnumformat{
     \ifblank\chapternumstyle\then\else\numstyle\chapternum.\fi
     \numstyle\sectionnum.}

\subsectionstyle{left}                           
\subsectionnumstyle{blank}                       
\def\subsectionbreak{\vskip0pt plus4\baselineskip\penalty-100
     \vskip0pt plus-4\baselineskip}              
\abovesubsectionskip=\the\baselineskip           
     plus.25\baselineskip minus.25\baselineskip
\belowsubsectionskip=.75\baselineskip            
     plus.19\baselineskip minus.19\baselineskip
\def\subsectionnumformat{
     \ifblank\chapternumstyle\then\else\numstyle\chapternum.\fi
     \ifblank\sectionnumstyle\then\else\numstyle\sectionnum.\fi
     \numstyle\subsectionnum.}


\def\undefinedlabelformat{$\bullet$}             


\equationnumstyle{arabic}                        
\subequationnumstyle{blank}                      
\figurenumstyle{arabic}                          
\subfigurenumstyle{blank}                        
\tablenumstyle{arabic}                           
\subtablenumstyle{blank}                         
\defnumstyle{arabic}                             
\subdefnumstyle{blank}                           
\thmnumstyle{arabic}                             
\subthmnumstyle{blank}                           
\lemnumstyle{arabic}                             
\sublemnumstyle{blank}                           

\eqnseriesstyle{alphabetic}                      
\figseriesstyle{alphabetic}                      
\tblseriesstyle{alphabetic}                      
\defseriesstyle{alphabetic}                      
\thmseriesstyle{alphabetic}                      
\lemseriesstyle{alphabetic}                      

\def\puteqnformat{\hbox{
     \ifblank\chapternumstyle\then\else\numstyle\chapternum.\fi
     \ifblank\sectionnumstyle\then\else\numstyle\sectionnum.\fi
     \ifblank\subsectionnumstyle\then\else\numstyle\subsectionnum.\fi
     \numstyle\equationnum
     \numstyle\subequationnum}}
\def\putfigformat{\hbox{
     \ifblank\chapternumstyle\then\else\numstyle\chapternum.\fi
     \ifblank\sectionnumstyle\then\else\numstyle\sectionnum.\fi
     \ifblank\subsectionnumstyle\then\else\numstyle\subsectionnum.\fi
     \numstyle\figurenum
     \numstyle\subfigurenum}}
\def\puttblformat{\hbox{
     \ifblank\chapternumstyle\then\else\numstyle\chapternum.\fi
     \ifblank\sectionnumstyle\then\else\numstyle\sectionnum.\fi
     \ifblank\subsectionnumstyle\then\else\numstyle\subsectionnum.\fi
     \numstyle\tablenum
     \numstyle\subtablenum}}
\def\putdefformat{\hbox{
     \ifblank\chapternumstyle\then\else\numstyle\chapternum.\fi
     \ifblank\sectionnumstyle\then\else\numstyle\sectionnum.\fi
     \ifblank\subsectionnumstyle\then\else\numstyle\subsectionnum.\fi
     \numstyle\defnum
     \numstyle\subdefnum}}
\def\putthmformat{\hbox{
     \ifblank\chapternumstyle\then\else\numstyle\chapternum.\fi
     \ifblank\sectionnumstyle\then\else\numstyle\sectionnum.\fi
     \ifblank\subsectionnumstyle\then\else\numstyle\subsectionnum.\fi
     \numstyle\thmnum
     \numstyle\subthmnum}}
\def\putlemformat{\hbox{
     \ifblank\chapternumstyle\then\else\numstyle\chapternum.\fi
     \ifblank\sectionnumstyle\then\else\numstyle\sectionnum.\fi
     \ifblank\subsectionnumstyle\then\else\numstyle\subsectionnum.\fi
     \numstyle\lemnum
     \numstyle\sublemnum}}


\referencestyle{sequential}                      
\referencenumstyle{arabic}                       
\def\putrefformat{\numstyle\referencenum}        
\def\referencenumformat{\numstyle\referencenum.} 
\def\putreferenceformat{
     \everypar={\hangindent=1em \hangafter=1 }%
     \def\\{\hfil\break\null\hskip-1em \ignorespaces}%
     \leftskip=\refnumindent\parindent=0pt \interlinepenalty=1000 }


\def\fmtversion{2.6M (June 1992)}

\catcode`\@=12

\def\ref#1{(\puteqn{#1})}
\def\label#1{\eqno\eqnlabel{#1}}
\font\bigboldfont=cmbx10 scaled \magstep2
\def\displayhead#1{{\bigboldfont \leftline{#1}}
\vskip-10pt
\line{\hrulefill}}
\def\section#1{\ifblank\sectionnumstyle\then
          \else\newsectionnum=\next \fi
\displayhead{\ifblank\sectionnumstyle\then\else\sectionnumformat\ \fi#1}
     }
\def\appendix#1{\ifblank\sectionnumstyle\then
          \else\newsectionnum=\next \fi
\displayhead{Appendix
    \ifblank\sectionnumstyle\then\else\sectionnumformat\ \fi#1}
     }


\input psfig.sty
\font\linef=line10
\font\linew=linew10
\font\circle=lcircle10
\chardef\ArrowNum="1B
\chardef\RArrowNum="2D
\chardef\UArrowNum="36

\def\RArrow{\linew \RArrowNum}
\def\UArrow{\linew \UArrowNum}
\chardef\LLURlineNum="00
\chardef\LRULlineNum="40
\def\LLURline{\rlap{\kern0.08mm\lower0.08mm\hbox{\linew \LLURlineNum}
              }\kern-0.08mm\raise0.08mm\hbox{\linew \LLURlineNum\hskip0.16mm}}
\def\LRULline{\rlap{\kern0.08mm\raise0.08mm\hbox{\linew \LRULlineNum}
              }\kern-0.08mm\lower0.08mm\hbox{\linew \LRULlineNum\hskip0.16mm}}
\def\LLURlineS{{\linef \LLURlineNum}}
\def\LRULlineS{{\linef \LRULlineNum}}
\font\bigmath=cmsy10 scaled \magstep1
\def\bbullet{{\bigmath \char"0F}}
\def\bcirc{{\bigmath \char"0E}}
\font\smallmath=cmsy7
\def\SymbolA{{\smallmath \char"0F}}
\def\SymbolB{{\smallmath \char"0E}}
\def\verline#1#2#3{\rlap{\kern#1mm\raise#2mm
                   \hbox{\vrule height #3mm depth 0 pt}}}
\def\horline#1#2#3{\rlap{\kern#1mm\raise#2mm
                   \vbox{\hrule width #3mm depth 0 pt}}}
\def\dashit{\vbox{\hrule width 0.1mm depth 0 pt}\hskip 0.4 mm}
\def\Dashit{\vbox{\hrule width 0.4mm depth 0 pt}\hskip 0.6 mm}
\def\dashdiag{$\dashit\raise0.5mm\hbox{\dashit}\raise1mm\hbox{\dashit}
              \raise1.5mm\hbox{\dashit}\raise2mm\hbox{\dashit}$}
\def\ddashdiag{$\raise2mm\hbox{\dashit}\raise1.5mm\hbox{\dashit}
               \raise1mm\hbox{\dashit}\raise0.5mm\hbox{\dashit}\dashit$}
\def\DashhorlineFive#1#2{\rlap{\kern#1mm\raise#2mm
                   \hbox{$\Dashit\Dashit\Dashit\Dashit\Dashit$}}}
\def\dashhorlineTen#1#2{\rlap{\kern#1mm\raise#2mm
                   \hbox{$\dashit\dashit\dashit\dashit\dashit
                   \dashit\dashit\dashit\dashit\dashit
                   \dashit\dashit\dashit\dashit\dashit
                   \dashit\dashit\dashit\dashit\dashit$}}}
\def\dashdiaglineTen#1#2{\rlap{\kern#1mm\raise#2mm
                   \hbox{$\hbox{\dashdiag}\raise2.5mm\hbox{\dashdiag}
                   \raise5mm\hbox{\dashdiag}\raise7.5mm\hbox{\dashdiag}$}}}
\def\ddashdiaglineTen#1#2{\rlap{\kern#1mm\raise#2mm
                   \hbox{$\raise7.5mm\hbox{\ddashdiag}\raise5mm\hbox{\ddashdiag}
                   \raise2.5mm\hbox{\ddashdiag}\hbox{\ddashdiag}$}}}
\def\dashdiaglineSevenFive#1#2{\rlap{\kern#1mm\raise#2mm
                   \hbox{$\hbox{\dashdiag}\raise2.5mm\hbox{\dashdiag}
                   \raise5mm\hbox{\dashdiag}$}}}
\def\ddashdiaglineSevenFive#1#2{\rlap{\kern#1mm\raise#2mm
                   \hbox{$\raise5mm\hbox{\ddashdiag}
                   \raise2.5mm\hbox{\ddashdiag}\hbox{\ddashdiag}$}}}
\def\Verline#1#2#3{\rlap{\kern#1mm\raise#2mm
                   \hbox{\vrule height #3mm width 0.7pt depth 0 pt}}}
\def\Horline#1#2#3{\rlap{\kern#1mm\raise#2mm
                   \vbox{\hrule height 0.7pt width #3mm depth 0 pt}}}
\def\putbox#1#2#3{\setbox117=\hbox{#3}
                  \dimen121=#1mm
                  \dimen122=#2mm
                  \dimen123=\wd117
                  \dimen124=\ht117
                  \divide\dimen123 by -2
                  \divide\dimen124 by -2
                  \advance\dimen121 by \dimen123
                  \advance\dimen122 by \dimen124
                  \rlap{\kern\dimen121\raise\dimen122\hbox{#3}}}
\def\LDArrow#1#2{\putbox{#1}{#2}{\linew\char"09}}
\def\RUArrow#1#2{\putbox{#1}{#2}{\linew\char"12}}
\def\RDArrow#1#2{\putbox{#1}{#2}{\linew\char"52}}
\def\LUArrow#1#2{\putbox{#1}{#2}{\linew\char"49}}
\def\offsetputbox#1#2#3#4#5#6{\setbox117=\hbox{#3}
                  \dimen121=#1mm
                  \dimen122=#2mm
                  \dimen123=\wd117
                  \dimen124=\ht117
                  \divide\dimen123 by -2
                  \divide\dimen124 by -2
                  \advance\dimen121 by \dimen123
                  \advance\dimen122 by \dimen124
                  \setbox117=\hbox{#4}
                  \dimen123=\wd117
                  \dimen124=\ht117
                  \multiply\dimen123 by #5
                  \divide\dimen123 by 100
                  \multiply\dimen124 by #6
                  \divide\dimen124 by 100
                  \advance\dimen121 by \dimen123
                  \advance\dimen122 by \dimen124
                  \rlap{\kern\dimen121\raise\dimen122\hbox{#3}}}
\def\leftputbox#1#2#3{\setbox117=\hbox{#3}
                  \dimen122=#2mm
                  \dimen124=\ht117
                  \divide\dimen124 by -2
                  \advance\dimen122 by \dimen124
                  \rlap{\kern#1mm\raise\dimen122\hbox{#3}}}
\def\rightputbox#1#2#3{\setbox117=\hbox{#3}
                  \dimen121=#1mm
                  \dimen122=#2mm
                  \dimen123=\wd117
                  \dimen124=\ht117
                  \multiply\dimen123 by -1
                  \divide\dimen124 by -2
                  \advance\dimen121 by \dimen123
                  \advance\dimen122 by \dimen124
                  \rlap{\kern\dimen121\raise\dimen122\hbox{#3}}}
\def\topputbox#1#2#3{\setbox117=\hbox{#3}
                  \dimen121=#1mm
                  \dimen123=\wd117
                  \divide\dimen123 by -2
                  \advance\dimen121 by \dimen123
                  \rlap{\kern\dimen121\raise#2mm\hbox{#3}}}
\def\lhvertex#1#2{\setbox117=\hbox{\LLURlineS}
                \dimen121=#1mm
                \dimen122=#2mm
                \dimen123=\wd117
                \dimen124=\ht117
                \rlap{\kern\dimen121\raise\dimen122\hbox{\LLURlineS}}
                \setbox117=\hbox{\LRULlineS}
                \dimen121=#1mm
                \dimen122=#2mm
                \dimen123=\wd117
                \dimen124=\ht117
                \multiply\dimen124 by -1
                \advance\dimen122 by \dimen124
                \rlap{\kern\dimen121\raise\dimen122\hbox{\LRULlineS}}}
\def\rhvertex#1#2{\setbox117=\hbox{\LLURlineS}
                \dimen121=#1mm
                \dimen122=#2mm
                \dimen123=\wd117
                \dimen124=\ht117
                \multiply\dimen123 by -1
                \multiply\dimen124 by -1
                \advance\dimen121 by \dimen123
                \advance\dimen122 by \dimen124
                \rlap{\kern\dimen121\raise\dimen122\hbox{\LLURlineS}}
                \setbox117=\hbox{\LRULlineS}
                \dimen121=#1mm
                \dimen122=#2mm
                \dimen123=\wd117
                \dimen124=\ht117
                \multiply\dimen123 by -1
                \advance\dimen121 by \dimen123
                \rlap{\kern\dimen121\raise\dimen122\hbox{\LRULlineS}}}
\def\vertex#1#2{\lhvertex{#1}{#2}\rhvertex{#1}{#2}}
\def\Blhvertex#1#2{\setbox117=\hbox{\LLURline}
                \dimen121=#1mm
                \dimen122=#2mm
                \dimen123=\wd117
                \dimen124=\ht117
                \rlap{\kern\dimen121\raise\dimen122\hbox{\LLURline}}
                \setbox117=\hbox{\LRULline}
                \dimen121=#1mm
                \dimen122=#2mm
                \dimen123=\wd117
                \dimen124=\ht117
                \multiply\dimen124 by -1
                \advance\dimen122 by \dimen124
                \rlap{\kern\dimen121\raise\dimen122\hbox{\LRULline}}}
\def\Brhvertex#1#2{\setbox117=\hbox{\LLURline}
                \dimen121=#1mm
                \dimen122=#2mm
                \dimen123=\wd117
                \dimen124=\ht117
                \multiply\dimen123 by -1
                \multiply\dimen124 by -1
                \advance\dimen121 by \dimen123
                \advance\dimen122 by \dimen124
                \rlap{\kern\dimen121\raise\dimen122\hbox{\LLURline}}
                \setbox117=\hbox{\LRULline}
                \dimen121=#1mm
                \dimen122=#2mm
                \dimen123=\wd117
                \dimen124=\ht117
                \multiply\dimen123 by -1
                \advance\dimen121 by \dimen123
                \rlap{\kern\dimen121\raise\dimen122\hbox{\LRULline}}}
\def\Bvertex#1#2{\Blhvertex{#1}{#2}\Brhvertex{#1}{#2}}
\def\dline#1#2{\putbox{#1}{#2}{\LLURlineS}}
\def\ddline#1#2{\putbox{#1}{#2}{\LRULlineS}}
\def\HalfCircle#1#2{\setbox117=\hbox{\circle\char"27}
                  \dimen121=#1mm
                  \dimen122=#2mm
                  \dimen123=\wd117
                  \dimen124=\ht117
                  \divide\dimen123 by -2
                  \multiply\dimen124 by -1
                  \advance\dimen121 by \dimen123
                  \advance\dimen122 by \dimen124
                  \rlap{\kern\dimen121\raise\dimen122
     \hbox{\offinterlineskip
     \vbox{\hbox{\circle \char"27}\hbox{\circle \char"26}}}}}

\def\plot#1#2{
  \putbox{#1}{#2}{$\cdot$}}

\def\diamond#1#2{
  \putbox{#1}{#2}{\SymbolB}}
\def\square#1#2{
  \putbox{#1}{#2}{\SymbolA}}

\chapternumstyle{blank}                           
\sectionnumstyle{arabic}                          
\def\cite#1{$\lbrack#1\rbrack$}
\def\bibitem#1{\parindent=10mm\item{\hbox to 8 mm{\cite{#1}\hfill}}}
\def\DDM{1}
\def\DoSch{2}
\def\DEHP{3}
\def\DeEv{4}
\def\HaNa{5}
\def\KSZa{6}
\def\KSZb{7}
\def\HiPeSa{8}
\def\Hinrichs{9}
\def\PeRy{10}
\def\Rujan{11}
\def\Baxter{12}
\def\BaLe{13}
\def\RSS{14}
\def\KaDoNi{15}
\def\Schuetz{16}
\def\OwBa{17}
\def\Sandow{18}
\def\EssRi{19}
\def\ADHR{20}
\def\KoelnB{21}
\def\Ma{22}
\def\GoVe{23}
\def\InKo{24}
\def\BaYu{25}
\def\Giac{26}
\def\HiKrPe{27}
\def\SchuetzBl{28}
\def\HiPriv{29}
\def\SchueSti{30}
\def\figindents{\leftskip=4.5 true pc \rightskip=4 true pc}
\setbox22 = \hbox{1}
\def\id{\rlap{1}\rlap{\kern 1pt \vbox{\hrule width 4pt depth 0 pt}}
        \rlap{\kern 3.5 pt \hbox{\vrule height \ht22 depth 0 pt}}
            \hskip\wd22}

\def\sn{\smallskip\noindent}
\def\mn{\medskip\noindent}
\def\bn{\bigskip\noindent}
\def\state#1{\vert #1 \rangle}
\def\astate#1{\langle #1 \vert}
\def\R{{\cal R}}
\def\L{{\cal L}}
\def\T{{\cal T}}
\def\Order{{\cal O}}
\def\frac#1#2{{#1 \over #2}}
\def\paramA{{\cal A}}
\def\abbrevB{{\cal B}}
\def\gt{\widetilde{g}}
\def\opA{A}
\def\opB{B}
\def\opAh{\widehat{A}}
\def\opBh{\widehat{B}}
\def\coA{a}
\def\coB{b}
\def\coAh{\widehat{a}}
\def\coBh{\widehat{b}}
\def\Ch{\widehat{C}}
\def\Jop{\widehat{J}}
\def\ph{\widehat{p}}
\def\qh{\widehat{q}}
\def\alphah{\widehat{\alpha}}
\def\betah{\widehat{\beta}}
\def\gammah{\widehat{\gamma}}
\def\deltah{\widehat{\delta}}
\def\kappah{\widehat{\kappa}}
\def\kappat{{\textstyle \widetilde{\kappa}}}

\def\abs#1{\vert #1 \vert}

\def\FD{F^{\dagger}}
\def\href#1#2{{#2}}
%
%
\font\large=cmbx10 scaled \magstep3
\font\bigf=cmr10 scaled \magstep2
\pageno=0
\def\folio{
\ifnum\pageno<1 \footline{\hfil} \else\number\pageno \fi}
\rightline{cond-mat/9606053}
\rightline{June 1996}
\vskip 2.0truecm
\centerline{\large Matrix-Product States for a One-Dimensional}
\vskip 0.5truecm
\centerline{\large Lattice Gas with Parallel Dynamics}
\vskip 1.5truecm
\centerline{\bigf A.\ Honecker and I.\ Peschel}
\bigskip\medskip
\centerline{\it Fachbereich Physik, Freie Universit\"at Berlin,}
\centerline{\it Arnimallee 14, D--14195 Berlin, Germany}
\vskip 1.9truecm
\centerline{\bf Abstract}
\vskip 0.2truecm
\noindent
The hopping motion of classical particles on a chain coupled
to reservoirs at both ends is studied for parallel dynamics
with arbitrary probabilities. The stationary state is obtained in the
form of an alternating matrix product. The properties of one-
and two-dimensional representations are studied in detail and a
general relation of the matrix algebra to that of the sequential
limit is found. In this way the general phase diagram of the
model is obtained. The mechanism of the sequential limit, the
formulation as a vertex model and other aspects are discussed.
\vfill
\noindent
{\bf Keywords:} Kinetic models, parallel dynamics, boundary effects,
                spin chains, vertex models, matrix-product states
\mn
\leftline{\hbox to 5 true cm{\hrulefill}}
\leftline{e-mail:}
\leftline{\quad honecker@omega.physik.fu-berlin.de}
\leftline{\quad peschel@aster.physik.fu-berlin.de}
\eject
\section{Introduction}
\mn
The study of classical kinetic models in one dimension has revealed
interesting physical properties (e.g.\ non-equilibrium phase transitions)
and mathematical structures. Moreover, one finds close connections with
quantum-mechanical spin problems. A prominent example is the diffusion
of particles with hard-core repulsion on the sites of a chain which is
coupled to reservoirs at both ends \cite{\DDM-\DeEv}. This model shows at
least three phases which differ in their density profiles and the current
through the system. Boundary effects, i.e.\ the rates at which particles
enter and leave the system, play an essential r\^ole. Mathematically,
the model can be described as a spin one-half problem and, if the
dynamics consists of single particle processes, the time evolution
operator of the master equation has the form of the Hamiltonian
of the Heisenberg model with boundary fields at both ends.
The stationary state can be written in the form of a matrix product,
where the weight of a configuration is given by an expression of
the form $\astate{W} A B B A B \ldots \state{V}$ with operators $A$, $B$
and vectors $\astate{W}$, $\state{V}$ in an auxiliary space. Such states,
which generalize simple product states so as to give non-trivial correlations,
were first found in a problem of lattice animals \cite{\HaNa} and
for certain quantum spin chains \cite{\KSZa,\KSZb}. They have
also been encountered in diffusion-coagulation models \cite{\HiPeSa}.
The detailed mechanism is somewhat different in each of the three
cases.
\mn
In the following we will study this model with a more general type
of dynamics. The time is taken discrete and in each time step,
hopping processes between half of all pairs of nearest-neighbour
sites can take place. This is not yet full parallel dynamics as
desired e.g.\ in traffic-flow problems, but we will still term it
`parallel'. This model has been considered before in the case
of deterministic uni-directional motion on the chain \cite{\Hinrichs}.
Here we will treat the general case where the particles hop with
probabilities $p$ and $q$ to the right and left, respectively. This contains the
deterministic as well as the sequential dynamics of the master
equation as special cases. The latter is obtained if all probabilities
tend towards zero. This limiting case corresponds to the Hamiltonian 
limit in two-dimensional statistical physics. In fact, there is
a close relation to that area, since the parallel dynamics can
be formulated as an asymmetric six-vertex model with additional
boundary terms.
\mn
It turns out that this general model has properties which are quite
similar to those of the sequential limit. One physical distinction
is a stationary density which alternates from site to site
and which results from the boundary terms combined with the
structure of our parallel dynamics. The matrix-product
groundstate, which also exists here, has a corresponding
sublattice structure. The mechanism for the stationary state is
the same as already encountered in the deterministic limit $p=1$,
$q=0$ \cite{\Hinrichs}. In each time step (corresponding to
the action of one row of vertices in the vertex model), the
sublattices exchange their r\^ole, so that after two steps the
state is reproduced. It should be mentioned that simple (scalar)
product states in two-dimensional models were studied already
some time ago. For example, a homogeneous state of that type
was found for special cases of eight-vertex models with
fields \cite{\PeRy}. Alternating states were considered in
\cite{\Rujan,\Baxter} for the case of IRF models which are the dual 
formulation of vertex models. In the context of spin systems,
such states are called `disorder solutions' and usually
result from competing interactions (see \cite{\BaLe} and
references therein).
\mn
To obtain the stationary state, one may try to find a
finite-dimensional representation of the four operators defining
the algebra and the corresponding vectors satisfying the boundary
relations. Here we present a scalar product state (i.e.\
a one-dimensional representation) and a representation in
terms of two-dimensional matrices. These representations
exist for special submanifolds in the parameter space. The
general situation, however, is more easily handled if one
notes that there exist representations where the quantities
on one sublattice may be eliminated such that one can
lift any representation of the algebra in the sequential
limit with suitably `renormalized' parameters to a representation
of the algebra for parallel dynamics. This feature has
also been discovered independently in the context of a dynamics
where the updating is done step-by-step from one end of the
chain to the other \cite{\RSS} \footnote{${}^{1})$}{
In this paper we will use a different terminology than in
\cite{\RSS}. We reserve the term `sequential' for a situation
where in each time step only one local process can take place,
but at a random position.
}.
Using this connection, we can take over results and methods
for the sequential case, e.g.\ to calculate the current
or the density profile in the general case. In this way,
one obtains a rather complete picture.
\mn
The paper is organized as follows. In Section 2 we
introduce the model, discuss the vertex formulation,
the Hamiltonian limit, and set up relations for the
matrix-product state. In Section 3 we find one- and
two-dimensional matrix representations for the algebra  
arising from the matrix-product ansatz which we use
in Section 4 to compute various physical quantities.
The mapping to the sequential case is described in Section 5
and used in Section 6 to compute the current for general
values of the parameters.
Section 7 contains a discussion of the results and remarks
on some other aspects as the relation to the model in
\cite{\RSS} and the integrability of the model.
Several appendices contain some supplementary material.
\bn
\section{Model and matrix-product ansatz}
\mn
The diffusion of particles with hard-core repulsion is a stochastic
process on a lattice which we here choose to be one-dimensional
with $N$ sites where $N$ is even. Each site can have two
states: It can either be empty or it can be occupied by
one particle. Particles can hop along the bonds of the lattice
onto empty sites, to the right with probability $p$ and
to the left with probability $q$. At the left and right
boundaries particles can be added or extracted. Particles
are added to an empty leftmost site with probability $\alpha$
and removed from it with probability $\gamma$.
At the right boundary they are extracted with probability $\beta$
and added with probability $\delta$. All theses processes can take
place simultaneously, but updates are performed in two steps in order
to permit at most one hopping process at each site in any time step.
Particles are removed and added at the boundaries during the first
time step. During this time step they can also hop along
the bonds connecting the even sites with the odd ones to their right.
During the second time step they can hop only along the other
bonds.  Because of the restriction that a site may be occupied
by at most one particle, this is a non-trivial many-body problem.
\mn
For these update rules the time transfer matrix $T$ which
describes the time evolution of the probability distribution
has the structure
$T = T_2 T_1$ where $T_1$ accounts for all processes that can
take place during the first time step, and $T_2$ for those
of the second one. They are given by
$$\eqalign{
T_1 &= \L \otimes \underbrace{\T \otimes \cdots \otimes \T}_{N-1 \ {\rm times}}
          \otimes \R \, , \cr
T_2 &= \underbrace{\T \otimes \T \otimes \cdots \otimes \T}_{N \ {\rm times}}
          \, , \cr
}\label{transMat}$$
where the matrices $\T$, $\L$ and $\R$ describe hopping and
particle input and output, respectively. In a suitable basis they
are given by
$$\T = \pmatrix{1&0&0&0 \cr
                0 & 1 - q & p & 0 \cr
                0 & q & 1 - p & 0 \cr
                0&0&0&1\cr} \, , \qquad
\L = \pmatrix{1-\alpha & \gamma \cr
              \alpha & 1-\gamma \cr} \, , \qquad
\R = \pmatrix{1-\delta & \beta \cr
              \delta & 1 - \beta \cr} \, .
\label{localTransMat}$$
This stochastic model can be regarded as a vertex model with the
time evolution operator $T$ corresponding to the diagonal-to-diagonal
transfer matrix \cite{\KaDoNi,\Schuetz}. This is shown in Fig.\ 1
where the particles sit on the bonds of the lattice and time evolves
upwards. The initial state before application of $T$ is given by the
configuration of the bonds at the lower edge of the shaded
region, and the final state is given by that at its upper border.
\mn
\mn
\mn
$$
\lhvertex{20}{3}
\vertex{34}{3}
\vertex{48}{3}
\vertex{62}{3}
\vertex{76}{3}
\rhvertex{90}{3}
\vertex{27}{10}
\vertex{41}{10}
\vertex{55}{10}
\vertex{69}{10}
\vertex{83}{10}
\Blhvertex{20}{17}
\Bvertex{34}{17}
\Bvertex{48}{17}
\Bvertex{62}{17}
\Bvertex{76}{17}
\Brhvertex{90}{17}
\Bvertex{27}{24}
\Bvertex{41}{24}
\Bvertex{55}{24}
\Bvertex{69}{24}
\Bvertex{83}{24}
\lhvertex{20}{31}
\vertex{34}{31}
\vertex{48}{31}
\vertex{62}{31}
\vertex{76}{31}
\rhvertex{90}{31}
\putbox{23}{12}{$\scriptstyle 1$}
\putbox{31}{12}{$\scriptstyle 2$}
\putbox{37}{12}{$\scriptstyle 3$}
\putbox{45}{12}{$\scriptstyle 4$}
\putbox{62}{12}{$\cdots$}
\putbox{77.5}{12}{$\scriptstyle N-1$}
\putbox{87}{12}{$\scriptstyle N$}
\putbox{24}{29}{$\scriptstyle 1$}
\putbox{30}{29}{$\scriptstyle 2$}
\putbox{38}{29}{$\scriptstyle 3$}
\putbox{44}{29}{$\scriptstyle 4$}
\putbox{69}{29}{$\cdots$}
\putbox{86}{29}{$\scriptstyle N$}
\dashhorlineTen{15}{13.5}
\dashhorlineTen{25}{13.5}
\dashhorlineTen{35}{13.5}
\dashhorlineTen{45}{13.5}
\dashhorlineTen{55}{13.5}
\dashhorlineTen{65}{13.5}
\dashhorlineTen{75}{13.5}
\dashhorlineTen{85}{13.5}
\dashhorlineTen{15}{14}
\dashhorlineTen{25}{14}
\dashhorlineTen{35}{14}
\dashhorlineTen{45}{14}
\dashhorlineTen{55}{14}
\dashhorlineTen{65}{14}
\dashhorlineTen{75}{14}
\dashhorlineTen{85}{14}
\dashhorlineTen{15}{14.5}
\dashhorlineTen{25}{14.5}
\dashhorlineTen{35}{14.5}
\dashhorlineTen{45}{14.5}
\dashhorlineTen{55}{14.5}
\dashhorlineTen{65}{14.5}
\dashhorlineTen{75}{14.5}
\dashhorlineTen{85}{14.5}
\dashhorlineTen{15}{15}
\dashhorlineTen{25}{15}
\dashhorlineTen{35}{15}
\dashhorlineTen{45}{15}
\dashhorlineTen{55}{15}
\dashhorlineTen{65}{15}
\dashhorlineTen{75}{15}
\dashhorlineTen{85}{15}
\dashhorlineTen{15}{15.5}
\dashhorlineTen{25}{15.5}
\dashhorlineTen{35}{15.5}
\dashhorlineTen{45}{15.5}
\dashhorlineTen{55}{15.5}
\dashhorlineTen{65}{15.5}
\dashhorlineTen{75}{15.5}
\dashhorlineTen{85}{15.5}
\dashhorlineTen{15}{16}
\dashhorlineTen{25}{16}
\dashhorlineTen{35}{16}
\dashhorlineTen{45}{16}
\dashhorlineTen{55}{16}
\dashhorlineTen{65}{16}
\dashhorlineTen{75}{16}
\dashhorlineTen{85}{16}
\dashhorlineTen{15}{16.5}
\dashhorlineTen{25}{16.5}
\dashhorlineTen{35}{16.5}
\dashhorlineTen{45}{16.5}
\dashhorlineTen{55}{16.5}
\dashhorlineTen{65}{16.5}
\dashhorlineTen{75}{16.5}
\dashhorlineTen{85}{16.5}
\dashhorlineTen{15}{17}
\dashhorlineTen{25}{17}
\dashhorlineTen{35}{17}
\dashhorlineTen{45}{17}
\dashhorlineTen{55}{17}
\dashhorlineTen{65}{17}
\dashhorlineTen{75}{17}
\dashhorlineTen{85}{17}
\dashhorlineTen{15}{17.5}
\dashhorlineTen{25}{17.5}
\dashhorlineTen{35}{17.5}
\dashhorlineTen{45}{17.5}
\dashhorlineTen{55}{17.5}
\dashhorlineTen{65}{17.5}
\dashhorlineTen{75}{17.5}
\dashhorlineTen{85}{17.5}
\dashhorlineTen{15}{18}
\dashhorlineTen{25}{18}
\dashhorlineTen{35}{18}
\dashhorlineTen{45}{18}
\dashhorlineTen{55}{18}
\dashhorlineTen{65}{18}
\dashhorlineTen{75}{18}
\dashhorlineTen{85}{18}
\dashhorlineTen{15}{18.5}
\dashhorlineTen{25}{18.5}
\dashhorlineTen{35}{18.5}
\dashhorlineTen{45}{18.5}
\dashhorlineTen{55}{18.5}
\dashhorlineTen{65}{18.5}
\dashhorlineTen{75}{18.5}
\dashhorlineTen{85}{18.5}
\dashhorlineTen{15}{19}
\dashhorlineTen{25}{19}
\dashhorlineTen{35}{19}
\dashhorlineTen{45}{19}
\dashhorlineTen{55}{19}
\dashhorlineTen{65}{19}
\dashhorlineTen{75}{19}
\dashhorlineTen{85}{19}
\dashhorlineTen{15}{19.5}
\dashhorlineTen{25}{19.5}
\dashhorlineTen{35}{19.5}
\dashhorlineTen{45}{19.5}
\dashhorlineTen{55}{19.5}
\dashhorlineTen{65}{19.5}
\dashhorlineTen{75}{19.5}
\dashhorlineTen{85}{19.5}
\dashhorlineTen{15}{20}
\dashhorlineTen{25}{20}
\dashhorlineTen{35}{20}
\dashhorlineTen{45}{20}
\dashhorlineTen{55}{20}
\dashhorlineTen{65}{20}
\dashhorlineTen{75}{20}
\dashhorlineTen{85}{20}
\dashhorlineTen{15}{20.5}
\dashhorlineTen{25}{20.5}
\dashhorlineTen{35}{20.5}
\dashhorlineTen{45}{20.5}
\dashhorlineTen{55}{20.5}
\dashhorlineTen{65}{20.5}
\dashhorlineTen{75}{20.5}
\dashhorlineTen{85}{20.5}
\dashhorlineTen{15}{21}
\dashhorlineTen{25}{21}
\dashhorlineTen{35}{21}
\dashhorlineTen{45}{21}
\dashhorlineTen{55}{21}
\dashhorlineTen{65}{21}
\dashhorlineTen{75}{21}
\dashhorlineTen{85}{21}
\dashhorlineTen{15}{21.5}
\dashhorlineTen{25}{21.5}
\dashhorlineTen{35}{21.5}
\dashhorlineTen{45}{21.5}
\dashhorlineTen{55}{21.5}
\dashhorlineTen{65}{21.5}
\dashhorlineTen{75}{21.5}
\dashhorlineTen{85}{21.5}
\dashhorlineTen{15}{22}
\dashhorlineTen{25}{22}
\dashhorlineTen{35}{22}
\dashhorlineTen{45}{22}
\dashhorlineTen{55}{22}
\dashhorlineTen{65}{22}
\dashhorlineTen{75}{22}
\dashhorlineTen{85}{22}
\dashhorlineTen{15}{22.5}
\dashhorlineTen{25}{22.5}
\dashhorlineTen{35}{22.5}
\dashhorlineTen{45}{22.5}
\dashhorlineTen{55}{22.5}
\dashhorlineTen{65}{22.5}
\dashhorlineTen{75}{22.5}
\dashhorlineTen{85}{22.5}
\dashhorlineTen{15}{23}
\dashhorlineTen{25}{23}
\dashhorlineTen{35}{23}
\dashhorlineTen{45}{23}
\dashhorlineTen{55}{23}
\dashhorlineTen{65}{23}
\dashhorlineTen{75}{23}
\dashhorlineTen{85}{23}
\dashhorlineTen{15}{23.5}
\dashhorlineTen{25}{23.5}
\dashhorlineTen{35}{23.5}
\dashhorlineTen{45}{23.5}
\dashhorlineTen{55}{23.5}
\dashhorlineTen{65}{23.5}
\dashhorlineTen{75}{23.5}
\dashhorlineTen{85}{23.5}
\dashhorlineTen{15}{24}
\dashhorlineTen{25}{24}
\dashhorlineTen{35}{24}
\dashhorlineTen{45}{24}
\dashhorlineTen{55}{24}
\dashhorlineTen{65}{24}
\dashhorlineTen{75}{24}
\dashhorlineTen{85}{24}
\dashhorlineTen{15}{24.5}
\dashhorlineTen{25}{24.5}
\dashhorlineTen{35}{24.5}
\dashhorlineTen{45}{24.5}
\dashhorlineTen{55}{24.5}
\dashhorlineTen{65}{24.5}
\dashhorlineTen{75}{24.5}
\dashhorlineTen{85}{24.5}
\dashhorlineTen{15}{25}
\dashhorlineTen{25}{25}
\dashhorlineTen{35}{25}
\dashhorlineTen{45}{25}
\dashhorlineTen{55}{25}
\dashhorlineTen{65}{25}
\dashhorlineTen{75}{25}
\dashhorlineTen{85}{25}
\dashhorlineTen{15}{25.5}
\dashhorlineTen{25}{25.5}
\dashhorlineTen{35}{25.5}
\dashhorlineTen{45}{25.5}
\dashhorlineTen{55}{25.5}
\dashhorlineTen{65}{25.5}
\dashhorlineTen{75}{25.5}
\dashhorlineTen{85}{25.5}
\dashhorlineTen{15}{26}
\dashhorlineTen{25}{26}
\dashhorlineTen{35}{26}
\dashhorlineTen{45}{26}
\dashhorlineTen{55}{26}
\dashhorlineTen{65}{26}
\dashhorlineTen{75}{26}
\dashhorlineTen{85}{26}
\dashhorlineTen{15}{26.5}
\dashhorlineTen{25}{26.5}
\dashhorlineTen{35}{26.5}
\dashhorlineTen{45}{26.5}
\dashhorlineTen{55}{26.5}
\dashhorlineTen{65}{26.5}
\dashhorlineTen{75}{26.5}
\dashhorlineTen{85}{26.5}
\dashhorlineTen{15}{27}
\dashhorlineTen{25}{27}
\dashhorlineTen{35}{27}
\dashhorlineTen{45}{27}
\dashhorlineTen{55}{27}
\dashhorlineTen{65}{27}
\dashhorlineTen{75}{27}
\dashhorlineTen{85}{27}
\dashhorlineTen{15}{27.5}
\dashhorlineTen{25}{27.5}
\dashhorlineTen{35}{27.5}
\dashhorlineTen{45}{27.5}
\dashhorlineTen{55}{27.5}
\dashhorlineTen{65}{27.5}
\dashhorlineTen{75}{27.5}
\dashhorlineTen{85}{27.5}
\HalfCircle{12}{20.5}
\putbox{10.5}{29.3}{\RArrow}
\putbox{0}{20.5}{$T$}
\hskip 110mm
$$
\sn

{\par\noindent\figindents
{\bf Fig.\ 1:} Representation of the hopping processes as a vertex
model. Only a part of the lattice is shown in the vertical
direction. The shaded region contains
all vertices that contribute to the diagonal-to-diagonal transfer
matrix $T$.
\par\noindent}
\mn
The local update operators \ref{localTransMat} can then be reinterpreted
in terms of the Boltzmann weights of all possible vertex configurations
as shown in Fig.\ 2 where the presence (absence) of a particle on a bond is
indicated by an arrow pointing upwards (downwards). This vertex model
is somewhat more complicated than the one treated in \cite{\OwBa}.
Usually, one considers the symmetric six-vertex model which is
invariant under inversion of the arrows (i.e.\ particle-hole
symmetry). For $p \ne q$ the bulk vertices in the first line
of Fig.\ 2 do not have this symmetry. More importantly, the
boundary vertices in the second line do not conserve the particle
number, because they correspond to particle injection and extraction.
\mn
\mn
$$
\vertex{50}{48}
\RUArrow{47.25}{45.25}
\RUArrow{51.5}{49.5}
\RDArrow{47.15}{51}
\RDArrow{51.4}{46.75}
\putbox{50}{40}{$p$}
\vertex{65}{48}
\LDArrow{63.75}{46.75}
\LUArrow{63.75}{49.4}
\LDArrow{68}{51}
\LUArrow{68}{45.15}
\putbox{65}{40}{$q$}
\vertex{80}{48}
\RUArrow{77.25}{45.25}
\LUArrow{78.75}{49.4}
\LDArrow{83}{51}
\RDArrow{81.4}{46.75}
\putbox{80}{40}{$1-p$}
\vertex{95}{48}
\LDArrow{93.75}{46.75}
\RUArrow{96.5}{49.5}
\RDArrow{92.15}{51}
\LUArrow{98}{45.15}
\putbox{95}{40}{$1-q$}
\vertex{35}{48}
\LDArrow{33.75}{46.75}
\LDArrow{38}{51}
\RDArrow{32.15}{51}
\RDArrow{36.4}{46.75}
\putbox{35}{40}{$1$}
\vertex{20}{48}
\RUArrow{17.25}{45.25}
\RUArrow{21.5}{49.5}
\LUArrow{18.75}{49.4}
\LUArrow{23}{45.15}
\putbox{20}{40}{$1$}
\lhvertex{33.25}{28}
\RUArrow{34.75}{29.5}
\RDArrow{34.65}{26.75}
\putbox{35}{20}{$\alpha$}
\rhvertex{51.75}{28}
\RUArrow{49}{25.25}
\RDArrow{48.9}{31}
\putbox{50}{20}{$\beta$}
\lhvertex{63.25}{28}
\LDArrow{66.25}{31}
\LUArrow{66.25}{25.15}
\putbox{65}{20}{$\gamma$}
\rhvertex{81.75}{28}
\LDArrow{80.5}{26.75}
\LUArrow{80.5}{29.4}
\putbox{80}{20}{$\delta$}
\lhvertex{33.25}{8}
\LDArrow{36.25}{11}
\RDArrow{34.65}{6.75}
\putbox{35}{0}{$1-\alpha$}
\rhvertex{51.75}{8}
\RUArrow{49}{5.25}
\LUArrow{50.5}{9.4}
\putbox{50}{0}{$1-\beta$}
\lhvertex{63.25}{8}
\RUArrow{64.75}{9.5}
\LUArrow{66.25}{5.15}
\putbox{65}{0}{$1-\gamma$}
\rhvertex{81.75}{8}
\LDArrow{80.5}{6.75}
\RDArrow{78.9}{11}
\putbox{80}{0}{$1-\delta$}
\hskip 115mm
$$
\sn

{\par\noindent\figindents
{\bf Fig.\ 2:} The Boltzmann weights for the vertex model
describing hopping with parallel dynamics. 
\par\noindent}
\mn
In this paper we shall be interested in the stationary state, i.e.\
a state that is invariant under the time evolution operator $T$.
In the language of the vertex model this corresponds to the
`groundstate' of the diagonal-to-diagonal transfer matrix.
\mn
Because of the sublattice structure of the
transfer matrix \ref{transMat} we make an alternating matrix-product
ansatz for the stationary state
$$\astate{W} \pmatrix{\opA \cr \opB \cr} \otimes
\pmatrix{\opAh \cr \opBh} \otimes \cdots \otimes
\pmatrix{\opA \cr \opB \cr} \otimes
\pmatrix{\opAh \cr \opBh} \state{V} \, ,
\label{matParAn}$$
where $\opA$, $\opB$, $\opAh$ and $\opBh$ are operators
in an auxiliary space. The operators $\opA$ and $\opAh$ describe
empty places while $\opB$ and $\opBh$ encode the presence of a
particle. $\state{V}$ and $\state{W}$ are vectors
in this auxiliary space which have to be chosen suitably and
have to satisfy the condition $\langle W \state{V} \ne 0$ in order
for \ref{matParAn} to be non-zero.
\mn
The mechanism of \cite{\Hinrichs} (and similarly \cite{\Baxter})
assumes that $T_1$ as well as $T_2$ exchange the operators $\opA$,
$\opB$ and $\opAh$, $\opBh$ with each other. At the boundaries this
gives rise to the conditions
$$\astate{W} \L
\pmatrix{\opA \cr \opB} =
\astate{W} \pmatrix{\opAh \cr \opBh} \, , \qquad
\R \pmatrix{\opAh \cr \opBh} \state{V}
= \pmatrix{\opA \cr \opB} \state{V}
\label{BoundParEq}$$
and for the interior one has
$$\T \left\{
\pmatrix{\opAh \cr \opBh} \otimes \pmatrix{\opA \cr \opB \cr} \right\}
= \pmatrix{\opA \cr \opB \cr} \otimes \pmatrix{\opAh \cr \opBh} \, .
\label{IntParEq}$$
After inserting the matrices \ref{localTransMat}, this ansatz leads
to 
\begineqnseries[parAlg]
$$\eqalignno{
\opAh \opA &= \opA \opAh \, , \qquad\quad
(1-q) \opAh \opB + p \opBh \opA = \opA \opBh \, , \cr
\opBh \opB &= \opB \opBh \, , \qquad\quad
q \opAh \opB + (1-p) \opBh \opA = \opB \opAh \, ,
&\eqnlabel{parAlgI} \cr
\{ (1-\delta) \opAh + \beta \opBh \} \state{V} &=
\opA \state{V} \, , \qquad
\{ \delta \opAh + (1-\beta) \opBh \} \state{V} =
\opB \state{V} \, , \cr
\astate{W} \{ (1-\alpha) \opA + \gamma \opB \} &=
\astate{W} \opAh \, , \qquad
\astate{W} \{ \alpha \opA + (1-\gamma) \opB \} =
\astate{W} \opBh \, .
&\eqnlabel{parAlgB} \cr
}$$
\endeqnseries
It has been argued in \cite{\Hinrichs} that the relations \ref{parAlg}
define a consistent associative algebra with Fock representation.
However, so far the problem
of finding the groundstate of $T$ has just been reformulated and
not yet been solved. In order to make further progress, one needs
a representation of the algebra defined by the $\opA$, $\opB$, $\opAh$
and $\opBh$ with suitable additional properties. This will be
the subject of Sections 3 and 5.
\mn
The corresponding matrix-product state of the sequential limit
is well-known (see e.g.\ \cite{\DEHP,\Sandow,\EssRi}). The sequential
limit is the limit of small probabilities, or equivalently the
Hamiltonian limit in the language of vertex models. In order to
be more precise set
$$x := \rho \, \widehat{x}
\label{ratesLimit}$$
for $x=p,q,\alpha,\beta,\gamma,\delta$
such that one can make an expansion in powers of $\rho$.
An immediate consequence of the definitions is that
$$\T = \id - \rho \, h \, , \qquad
\L = \id - \rho \, h_L \, , \qquad
\R = \id - \rho \, h_R \, ,
\label{localHam}$$
with matrices $h$, $h_L$ and $h_R$ that are independent of $\rho$
and describe the local processes with rates $\ph$, $\qh$, $\alphah$,
$\betah$, $\gammah$ and $\deltah$. The transfer matrix $T$
now takes the form
$$T = \id - \rho \, H + \Order(\rho^2) \, ,
\label{transMatLim}$$
where the Hamiltonian $H$ contains only nearest-neighbour interactions.
After a similarity transformation, this Hamiltonian becomes
the $U_q(su(2))$-invariant Hamiltonian of the ferromagnetic
XXZ-Heisenberg model \cite{\ADHR} with additional boundary terms $h_L$ and $h_R$
(see e.g.\ \cite{\EssRi}). Since this Hamiltonian does not have any sublattice
structure one can make a homogenous matrix-product ansatz. We set
$\opA = \opAh = E$ and $\opB = \opBh = D$ such that the operators with and
without hat in \ref{matParAn} become equal. Then one imposes the following
relations \cite{\DEHP,\Sandow,\EssRi} (see also \cite{\HiPeSa}) at the boundaries
$$\astate{W} h_L \pmatrix{E \cr D \cr} =
\astate{W} \pmatrix{-1 \cr 1 \cr} \, , \qquad
h_R \pmatrix{E \cr D \cr} \state{V}
= - \pmatrix{-1 \cr 1} \state{V} \, ,
\label{BoundHamEq}$$
and the following algebra for the bulk
$$h \left\{
\pmatrix{E \cr D\cr }
 \otimes \pmatrix{E \cr D\cr } \right\}
= \left\{
   \pmatrix{E \cr D \cr } \otimes
   \pmatrix{-1 \cr 1 \cr}
 - \pmatrix{-1 \cr 1 \cr} \otimes
   \pmatrix{E \cr D \cr } \right\} \, .
\label{IntHamEq}$$
With this ansatz the right term in \ref{IntHamEq} at site $x$
cancels the left term at site $x+1$ if the complete Hamiltonian
is applied. Finally, one is left with two boundary terms
which are cancelled by \ref{BoundHamEq}.
\mn
After inserting the explicit matrices $h$, $h_L$ and $h_R$
one finds that \ref{BoundHamEq} and \ref{IntHamEq} are equivalent to
$$\eqalign{
\ph D E - \qh E D &= D + E \, , \cr
\left( \betah D - \deltah E \right) \state{V} &= \state{V} \, , \cr
\astate{W} \left( \alphah E - \gammah D \right) &= \astate{W} \, . \cr
}\label{seqAlg}$$
This algebra can be used to compute
expectation values in the stationary state efficiently by means
of recurrence relations (see \cite{\DEHP,\Sandow,\EssRi}). The
operator $C := E + D$ acts like a transfer matrix in the
spatial direction. Obviously, $\astate{W} C^N \state{V}$ is
the sum of all coefficients of the groundstate. In order to
obtain a probability distribution one has to divide by this
factor. Then e.g.\ the density profile $\langle \tau_x \rangle$
is given by the expectation values $\langle \tau_x \rangle =
\astate{W} C^{x-1} D C^{N-x} \state{V} / \astate{W} C^N \state{V}$.
\mn
One of the main motivations for the work to be reported below was to find out
how the mechanism \ref{BoundParEq}, \ref{IntParEq} is related to the known
mechanism \ref{BoundHamEq}, \ref{IntHamEq} in the sequential limit.
In the next Section we will first see what the relation is in two
special cases.
\bn
\section{One- and two-dimensional representations}
\mn
First we consider a {\it scalar} product state. For this
representation of the algebra \ref{parAlg}
the operators $\opA$, $\opB$, $\opAh$ and $\opBh$ are
mapped to real numbers $\coA$, $\coB$, $\coAh$ and $\coBh$
and the boundary vectors can be discarded.
(We do not write the representation map explicitly and
below we will also use the same notation for the algebra
and its representation). After fixing the normalization
suitably, the condition \ref{parAlgB} is equivalent to
$$\eqalign{
\coA &= \delta \gamma - \gamma - \beta + \beta \gamma \, , \qquad
\coB = \delta \alpha - \alpha - \delta + \beta \alpha \, , \cr
\coAh &= \beta \gamma - \gamma - \beta + \beta \alpha \, , \qquad
\coBh = \delta \gamma - \alpha - \delta + \delta \alpha \, . \cr
}\label{repOneD}$$
The bulk relations \ref{parAlgI} are satisfied if
$(1-q) \coAh \coB - (1-p) \coBh \coA = 0$. Inserting the
result \ref{repOneD} one finds that the probabilities
$p$, $q$, $\alpha$, $\beta$, $\gamma$ and $\delta$ have to
obey the relation
$$\eqalign{
&(1-q) (\beta \gamma - \gamma - \beta + \beta \alpha)
      (\delta \alpha - \alpha - \delta + \beta \alpha) \cr
&= (1-p) (\delta \gamma - \alpha - \delta + \delta \alpha)
       (\delta \gamma - \gamma - \beta + \beta \gamma) \, . \cr
}\label{RelOneD}$$
Note that this condition is symmetric under the exchange of
$p$ with $q$, $\alpha$ with $\delta$ and $\beta$ with $\gamma$.
This symmetry is to be expected because parity (reflection
at the centre of the chain) effects precisely
this exchange of the probabilities. A further expansion of
\ref{RelOneD} is not very enlightning, but a few special cases
may be worth while mentioning.  For $\gamma = \delta = 0$ but
$\alpha \ne 0 \ne \beta$ the relation \ref{RelOneD} specializes
to
$$(1-q)(\alpha + \beta - \alpha \beta) = p - q \, .
\label{RelOneDtwoR}$$
This condition agrees with the one that has been found in
\cite{\RSS} for the existence of a one-dimensional representation
with $\gamma = \delta = 0$.
Another interesting case is the sequential limit. This is obtained
by inserting \ref{ratesLimit} into \ref{RelOneD}
and keeping only the leading non-trivial order in $\rho$.
As expected, one obtains a condition that is equivalent
to the known result for sequential dynamics, namely eq.\ (78)
in \cite{\EssRi}.
\mn
We would like to conclude the discussion of the one-dimensional
representation with a remark on more general mechanisms than
\ref{BoundParEq}, \ref{IntParEq}. Firstly, one can get rid of
the alternating structure in space by introducing a four-state
model that uses block-spin variables for the states on
any odd site and the even site to its right. Then one can make
a homogenous product state ansatz. If one further requires
the coefficients of this state to be independent of $N$ for
all even $N \ge 2$, one finds that the relation \ref{RelOneD}
must be satisfied. Even more strongly, one finds that the
block-spin product state can be factorized in the product state
that we have already found above. In this sense, the mechanism
\ref{BoundParEq}, \ref{IntParEq} gives rise to the most general
scalar product state.
\mn
Now we turn to the more interesting case of a {\it two-dimensional}
matrix-product state. As we have already indicated at the end
of the previous Section, the matrices 
$$C := \opA + \opB \, , \qquad \Ch := \opAh + \opBh \, ,
\label{defLocTM}$$
will play the r\^ole of transfer matrices in space. Therefore, it
is desirable to find representations where $C$ and $\Ch$ have a
particularly simple form. First one observes by summing over
the columns in \ref{IntParEq} that $[C, \Ch] = 0$. Since $C$
and $\Ch$ commute they can be simultaneously brought into
diagonal --or more generally-- Jordan normal form. If $C$ and $\Ch$ are
both invertible, an even more stronger statement holds
\cite{\Hinrichs}, namely that one can find an equivalent representation
where $C=\Ch$. On the other hand, one can check that any non-trivial
two-dimensional representation where $C \ne c \, \Ch$ with some
constant $c$ leads to the condition \ref{RelOneD} and is therefore
not interesting. For this reason we will from now on only consider
representations where
$$C = \Ch \, .
\label{localTmEq}$$
This choice also ensures that half of the boundary equations
\ref{BoundParEq} are satisfied automatically.
\mn
First we consider representations where $C = \Ch$ can be
diagonalized. We fix the normalization of the representation
by requiring that $C_{1,1} = 1$. So, apart from the matrix
elements e.g.\ of $\opA$ and $\opAh$ (and the vectors
$\state{V}$, $\state{W}$), only $C_{2,2}$ is still free.
With this choice we first solve the bulk equations \ref{parAlgI}.
One solution is given by
$$\eqalign{
C = \Ch =& \pmatrix{
1&0 \cr
0&-{\frac {\left (\paramA (p-q) (p+q-1) + p (1-p) \right )
\left (\paramA (p-q) -p\right )}
{\left (\paramA (p-q)-p +1 \right )pq}} \cr 
} \, , \cr
\opA =& \pmatrix{
-{\frac {\paramA \left (q-1\right)}{\paramA (p-q)-p +1}}&1 \cr
0&{\frac {\left (q-1\right )\left (\paramA (p-q) -p\right )\paramA }
{\left (\paramA (p-q)-p +1\right )p}} \cr
} \, , \cr
\opAh =& \pmatrix{
\paramA &1 \cr
0&{\frac {\paramA \left (\paramA (p-q) (p+q-1) + p (1-p) \right )}{
\left (\paramA (p-q)-p +1\right )p}} \cr
} \, , \cr
}\label{bulkRep2D}$$
where the constant $\paramA$ remains free. In obtaining
\ref{bulkRep2D} we have used the freedom to rescale the
two basis vectors in the representation space independently
to set $\opAh_{1,2} = 1$.
\mn
Now we have to solve the boundary equations \ref{parAlgB}.
Non-trivial solutions
for the vectors $\state{V}$ and $\state{W}$ exist iff
$$
\det\left( (1 - \alpha) \opA + \gamma \opB - \opAh\right) =
0 = \det\left( (1 - \delta) \opAh + \beta \opBh - \opA \right) \, .
\label{nonTrivBvec}$$
The solution of \ref{nonTrivBvec} can be reduced to a discussion of
the diagonal elements since all matrices in \ref{bulkRep2D} are
upper triangular. The relevant solution is given by the vanishing of the
(1,1)-element in the left matrix and the (2,2)-element of the right one. 
From this one first finds that the constant $\paramA$ is given by
$$\eqalign{
\paramA =&
\bigl(\left (1-p\right )\left (pq\gamma+q\delta\,\gamma\,\left (1-p
-q\right )+p\beta\,\left (p+\gamma\,\left (1-p-q\right )\right )
\right )\bigr) \cr
& \bigl(
q\delta\,\left (1-q\right )\left (q\left (1-\alpha\right )+
\alpha+\gamma\right )+p\beta\,\left (1-p\right )\left (p\left (1-\gamma
\right )+\alpha+\gamma\right ) + \cr
& pq\left (\left (1-2\,\delta-\beta
\right )\gamma+\left (1-2\,\beta-\delta\right )\alpha\right )+pq\left 
(\left (q\alpha+p\gamma\right )\left (\beta+\delta-1\right )+q\gamma\,
\delta+p\alpha\,\beta\right )\bigr)^{-1}
\, ,\cr
}\label{paramAval}$$
and with this result the following relation between the parameters
of the model:
$$\eqalignno{
&\left (1-p\right )\left (1-q\right )\left (
{p}^{3}{\alpha}^{2}{\beta}^{2}-{q}^{3}{\gamma}^{2}{\delta}^{2}\right ) &\cr
& +\left (1-p\right )p{q}^{2}
\gamma\delta\left (\alpha\beta\left (1-q\right )^{2}+\left (1
-q\right )\left( q \left (\alpha+\beta\right ) -\left (\delta
\alpha+\beta\gamma\right )\right)-q\left (\left (1-\gamma\right )\delta+
\gamma-q\right )\right ) &\cr
& -\left (1-q\right ){p}^{2}q\alpha\beta
\left (\gamma\delta\left (1-p\right )^{2}+\left (1-p\right )\left(p
\left (\gamma+\delta\right )-\left (\delta\alpha+
\beta\gamma\right )\right)-p\left (\left (1-\alpha\right )\beta+\alpha-p
\right )\right ) &\cr
&+{p}^{2}{q}^{2}\left (\alpha\,\gamma\,\left (\left (1-
\delta\right )^{2}-\left (1-\beta\right )^{2}\right )+\beta\,\delta
\,\left (\left (1-\gamma\right )^{2}-\left (1-\alpha\right )^{2}
\right )\right ) &\cr
&+{p}^{2}{q}^{3}\left (\beta\,\delta\,\left (1-\alpha
\right )^{2}+\alpha\,\gamma\,\left (1-\beta\right )^{2}-\gamma\,
\delta\,\left (1-\alpha\right )\left (1-\beta\right )\right ) &\cr
&-{p}^{3}{q}^{2}\left (\alpha\,\gamma\,\left (1-\delta\right )^{2}+\beta\,
\delta\,\left (1-\gamma\right )^{2}-\alpha\,\beta\,\left (1-\gamma
\right )\left (1-\delta\right )\right ) &\cr
& +{p}^{2}{q}^{2}\left ({q}^{2}
\gamma\,\delta\,\left (1-\alpha\right )\left (1-\beta\right )-{p}^{2
}\alpha\,\beta\,\left (1-\gamma\right )\left (1-\delta\right )
\right ) 
\qquad = 0 \, . &\eqnlabel{RelTwoD}\cr
}$$
This is the condition for the existence of a two-dimensional
matrix-product state.
Like \ref{RelOneD} this condition is symmetric under the exchange of
$p$ with $q$, $\alpha$ with $\delta$ and $\beta$ with $\gamma$.
\mn
To specify the two-dimensional representation completely one has
to determine the boundary vectors $\astate{W}$ and $\state{V}$
that belong to the solution \ref{bulkRep2D}, \ref{paramAval} and
\ref{RelTwoD}. They are straightforwardly obtained by determining the
nullspaces of the matrices in \ref{nonTrivBvec}: 
$$\eqalign{
\state{V} =&
\pmatrix{
\left(\beta+\delta\right)\left(\paramA (p-q)+1-p \right) \cr
{\paramA}^{2}\left (1-\delta-\beta\right )\left (p-q\right )-
\paramA\,\left (p-q+\left (1-p\right )\delta+\beta\,\left (1+q-2\,
p\right )\right )+\beta\,\left (1-p\right ) \cr
} \, , \cr
\state{W} =&
\pmatrix{
\left (1-\paramA\right )\left\{ \paramA \left (p-q\right )\left (p
q-\gamma p\left (1-p\right )-\alpha q\left (1-q\right )\right )+
\gamma {p}^{2}\left (1-p\right ) \right\}
-\paramA \alpha {q}^{2} \left (1-q\right ) \cr
pq \left(\alpha+ \gamma\right )\left(\paramA (p-q)+1-p\right) \cr
} . \cr
}\label{boundVecTwoD}$$
Inserting herein \ref{paramAval} one can check that $\langle W \state{V}$
is neither zero nor singular on the manifold given by \ref{RelTwoD}.
In passing we note that the case of symmetric diffusion in
the bulk of the chain ($q=p$) is not permitted because in
this case the normalization vanishes, i.e.\ $\langle W \state{V}=0$.
\mn
In order to illustrate the meaning of \ref{RelTwoD} we are now going
to look at a few special cases of it. Firstly, consider deterministic
hopping exclusively to the right, i.e.\ $p=1$ and $q=0$.
One observes that in this special case the condition \ref{RelTwoD}
is satisfied automatically, indicating that one can find a
two-dimensional representation for arbitrary boundary probabilities
$\alpha$, $\beta$, $\gamma$, $\delta$ if the bulk rates are
$p=1$, $q=0$. This is consistent with \cite{\Hinrichs} where
a two-dimensional representation was written down at
$p=1$, $q=\gamma=\delta = 0$ and arbitrary probabilities
$\alpha$ and $\beta$. However, it should be noted that we
may not directly set $p=1$ in our solution, because then one finds
$\paramA = 0$ from \ref{paramAval} and singularities
appear in \ref{bulkRep2D}.
\mn
Another interesting case is obtained when $\gamma = \delta = 0$,
but $p \ne 0$, $\alpha \ne 0 \ne \beta$. Here \ref{RelTwoD}
simplifies considerably:
$$(1 - q) \{(1-p-q) \alpha \beta + q (\alpha+\beta) \} = q (p-q) \, .
\label{RelTwoDtwoR}$$
This is a hyperbola in the $\alpha$-$\beta$-plane which intersects
the axes at $(p-q)/(1-q)$.
\mn
A final case of particular interest is the Hamiltonian limit.
This limit of the condition \ref{RelTwoD} is obtained by
inserting \ref{ratesLimit} and keeping only the leading order
in $\rho$. Like in the case of a one-dimensional representation
we recover the known result, namely eq.\ (81) of \cite{\EssRi}.
\mn
One might think that the condition \ref{RelTwoD} arises from the
special mechanism \ref{BoundParEq} and \ref{IntParEq}. However, 
for $\gamma = \delta = 0$ we have checked that the solution
obtained above is the most general two-dimensional one in the
following sense: As mentioned at the end of the discussion of
the scalar product state one eliminates the alternating structure
in space by introducing a four-state model that uses block-spin
variables for two neighbouring sites. Then one can make a
homogenous matrix-product ansatz with four independent operators
for each two-site block. One further requires the matrix
representation of the operators to be independent of the chain
length $N$. Then already a comparison with
the groundstate of the transfer matrix $T$ at $N \in \{2,4,6\}$
leads to the condition \ref{RelTwoDtwoR}.
\mn
There also exist representations where $C = \Ch$ is not
diagonalizable. We do not discuss them explicitly here
because we are not going to make use of them in the sequel.
The interested reader may find some results in appendix A.
We mention that non-trivial Jordan forms give rise to power
laws with positive exponents in the correlation functions
(see e.g.\ \cite{\HiPeSa}), in particular for two-dimensional
matrices to a density profile that is linear in space.
Physically, this is related to phase transitions in the system.
\mn
Finally, we turn to the question that was one of our motivations for
constructing the above matrix-product state, namely how
the state given by \ref{bulkRep2D}, \ref{paramAval}
-- \ref{boundVecTwoD} behaves in the sequential limit.  First one
observes that the matrices \ref{bulkRep2D} have the property
$$\opA - \opAh = \opBh - \opB =
{\paramA (1-\paramA) (p-q) \over 1-p+(p-q) \paramA} \id
= g(p,q;\alpha,\beta,\gamma,\delta) \id \, .
\label{matDiffId}$$
The function $g$ is obtained by inserting \ref{paramAval} and
too complicated to be presented here.
Obviously, an analogous results holds also for the one-dimensional
case because the difference of any two numbers trivially is a
multiple of the identity.
\mn
Having observed the property \ref{matDiffId} for the two-dimensional
representation, it is natural to restrict one's attention to those
representations of the algebra \ref{parAlg} where  
$\opA - \opAh$ and $\opBh - \opB$ are represented by the same
multiple of the identity. As we will see soon, such
a condition enables one to make full use of the machinery developed
for the algebra \ref{seqAlg} in \cite{\Sandow} and \cite{\EssRi}.
\mn
However, let us for the moment concentrate on the one- and
two-dimensional representations. Using \ref{matDiffId} to eliminate
e.g.\ $\opAh$ and $\opBh$, the conditions \ref{parAlg} turn into
$$\eqalign{
\astate{W} \left(\alpha \opA - \gamma \opB \right) &=
  g\, \astate{W} \, , \cr
\left(\beta \opB - \delta \opA \right) \state{V} &=
  g\, \left(1 - \beta - \delta\right) \state{V} \, , \cr
p \opB \opA - q \opA \opB &= g\,
    \left(((1-q) \opB + (1-p) \opA \right) \, . \cr
}\label{HamLimAlg}$$
This looks already very similar to \ref{seqAlg}. In order to
establish the correspondence manifestly, one inserts
\ref{ratesLimit} into the two-dimensional representation
\ref{bulkRep2D}, \ref{paramAval} --
\ref{boundVecTwoD} and also into \ref{HamLimAlg}.
Then it is straightforward to check that the limits
$\rho \to 0$ of all four operators $\opA$, $\opB$, $\opAh$,
$\opBh$ exist and are non-zero. Furthermore, one finds that
$\opAh \to  \opA$ and $\opBh \to \opB$ as is
to be expected because the alternating sublattice structure
must vanish in the sequential limit. The boundary vectors
with the normalization as in \ref{boundVecTwoD} have the property
$\state{V} = \rho \left(\state{V^{(0)}} + \Order(\rho) \right)$
and
$\state{W} = \rho^3 \left(\state{W^{(0)}} + \Order(\rho) \right)$
with non-vanishing vectors $\state{V^{(0)}}$ and $\state{W^{(0)}}$.
Finally, the function $g$ appearing in \ref{HamLimAlg} behaves
as $g(p,q;\alpha,\beta,\gamma,\delta) = \rho \,
\gt(\ph,\qh;\alphah,\betah,\gammah,\deltah) + \Order(\rho^2)$
with a non-vanishing function $\gt$ of the rates in the sequential
limit. Now one inserts all this into \ref{HamLimAlg}
and keeps only the leading orders in $\rho$. Identifying
finally $\lim_{\rho \to 0} \opA =: \gt \, E$
and $\lim_{\rho \to 0} \opB =: \gt \, D$
one recovers precisely \ref{seqAlg}. Thus, we have shown
for the one- and two-dimensional representations constructed above
that the matrix-product mechanism \ref{BoundParEq}, \ref{IntParEq}
leads precisely to the known mechanism \ref{BoundHamEq}, \ref{IntHamEq}
in the sequential limit. The crucial step to this end was the
observation \ref{matDiffId}.
\bn
\section{Physical quantities from the representations}
\mn
In this Section we present results for quantities
of physical interest such as the bulk density, the current and
the correlation length. The computations are based on
the one- and two-dimensional representation of the previous
Section. In order to present the results in
a compact form it will be convenient to introduce an
abbreviation which (as we will explain in more detail later)
is related to the variables $\kappa_{\pm}$
used in \cite{\Sandow,\EssRi}:
$$\eqalign{
\kappat_{\pm}(x,y) =
{1 \over 2 x} \Bigl( & -x(1-q) + y (1-p) + p - q \cr
& \pm \sqrt{
\left (-x(1-q) + y (1-p) + p - q \right )^2 + 4 x y
\left (1-q\right )\left (1-p\right ) }
\Bigr) \, , \cr
}\label{defKappaT}$$
where either $\alpha$ and $\gamma$ or $\beta$ and $\delta$
will be substituted for the arguments $x$ and $y$.
To understand the meaning of this quantity
consider the special case $\gamma=\delta=0$.
First one notes that $\kappat_{+}(x,0) = - (1-q)+(p - q) / x$
and $\kappat_{-}(x,0) = 0$. Thus, for $p \ne q$
and zero boundary probability $y$, the expression
$\kappat_{+}(x,0)$ is basically the inverse of the probability $x$
whereas $\kappat_{-}(x,0)$ is trivial. Turning on $y > 0$
means that one not only injects particles e.g.\ at the left
boundary but also extracts them again. Now the
variables $\kappat_{+}(x,y)$ can be thought of as
effective particle input and output rates.
They are non-trivial because particles can be e.g.\ injected
at the left boundary, then diffuse some time in the
interior before they return to the left boundary where
they are then extracted again. If this interpretation
of the $\kappat_{+}(x,y)$
is correct, they should be related to the ratio of $x$
and $y$ if hopping in the bulk is sufficiently suppressed.
Indeed, in the limit $p \to 0$, $q \to 0$ one finds
from \ref{defKappaT} that $\kappat_{+}(x,y) \to y/x$.
\mn
It should be noted that wherever the abbreviations $\kappat_{\pm}$
are used one has to be careful with the range of
validity of the result due to the ambiguity of the sign of the
square root in \ref{defKappaT}. In order to be on safe
grounds we restrict to $p > q$. The results for the
case $p < q$ can be recovered using parity, i.e.\ by exchanging
$p$ with $q$, $\alpha$ with $\delta$ and $\beta$ with $\gamma$.
\mn
Now we turn to the computation of physical quantities. We shall
be interested only in the thermodynamic limit $N \to \infty$.
In this limit, the computation amounts to taking appropriate
matrix elements of the relevant operators as the following
argument shows. Let $C_{i,i}$ be the largest of the diagonal
matrix elements of $C$. Then in $C^n \state{V}$
and in $\astate{W} C^m$ the $i$th basis vector in the auxiliary space
will yield the dominant contribution if $n$ respectively $m$
is large. Thus, only the $i$th matrix elements of the operators
corresponding to the quantities of interest will contribute in
the thermodynamic limit. Below, we will refrain from presenting
results that are obtained immediately by inserting the matrix
elements of the one- and two-dimensional representations,
but will instead present a formulation using the $\kappat_{\pm}$
that we have just introduced. These variables are used because
we believe that the results given in terms of them will
be valid also off the manifolds in the phase diagram accessible
by the one- and two-dimensional representations (see also
Section 6).
\mn
We first compute the density profile which is given by
$\langle\tau_x\rangle = \astate{W} C^{x-1} \opB C^{N-x} \state{V}/$
$\astate{W} C^N  \state{V}$ for $x$ odd and by
$\langle\widehat{\tau}_x\rangle = \astate{W} C^{x-1} \opBh C^{N-x} \state{V}/
 \astate{W} C^N  \state{V}$ for $x$ even.
These profiles are independent of $x$ for any scalar product
state. Their values can be written down immediately for the
product state \ref{repOneD}. The only non-trivial computation
is to show that the resulting expressions are equivalent to 
\begineqnseries[resDens1D2D]
$$\langle\tau_x\rangle =
{\kappat_{+}(\beta,\delta) \over 1 - q + \kappat_{+}(\beta,\delta)} \, , \qquad
\langle\widehat{\tau}_x\rangle =
{\kappat_{+}(\beta,\delta) \over 1 - p + \kappat_{+}(\beta,\delta)}
\label{resDens1D2Da}$$
and also to
$$
\langle\tau_x\rangle =
{1-p \over 1 - p + \kappat_{+}(\alpha,\gamma)} \, , \qquad
\langle\widehat{\tau}_x\rangle =
{1-q \over 1 - q + \kappat_{+}(\alpha,\gamma)}
\label{resDens1D2Db}$$
\endeqnseries
on the manifold given by \ref{RelOneD}.
\mn
For the two-dimensional representation \ref{bulkRep2D},
\ref{paramAval} and \ref{boundVecTwoD}, the density profiles
depend on the spatial variable $x$. However,
in the bulk of the system, the densities become constant and
follow easily from the observation made above:
$\langle\tau_x\rangle \approx \opB_{i,i}/C_{i,i}$ and
$\langle\widehat{\tau}_x\rangle \approx \opBh_{i,i}/C_{i,i}$ for
$1 \ll x \ll N$.
For the two-dimensional representation \ref{bulkRep2D}
and \ref{paramAval} two cases have be distinguished:
$C_{2,2} > C_{1,1}$ which we denote by `I' and $C_{1,1} > C_{2,2}$
which we denote by `II'. This corresponds to two different
regions in the phase diagram, see Fig.\ 3 below.
\mn
The results obtained from the two-dimensional
representation for the bulk ($1 \ll x \ll N$) can be identified with
the result \ref{resDens1D2D} on the manifold given by \ref{RelTwoD}.
One finds that \ref{resDens1D2Da} is valid in case I and
\ref{resDens1D2Db} applies to case II.
Note that we recover the result for sequential dynamics (eq.\ (100)
of \cite{\EssRi}) in the limit of small hopping probabilities.
Like in \cite{\EssRi} we regard the fact that we obtain the same result
from the one- and two-dimensional representations as an
indication that the expressions \ref{resDens1D2D} may be valid
throughout the corresponding phases. For the special case
$\gamma = \delta = 0$ they simplify
to
\begineqnseries[resDens1D2Dspec]
$$\eqalignno{
\langle\tau_x\rangle &= 1 - \beta {1-q \over p -q} \, , \qquad\
\langle\widehat{\tau}_x\rangle = {1 \over 1 -\beta} \langle\tau_x\rangle
\, , &\eqnlabel{resDens1D2Dspeca} \cr
\langle\tau_x\rangle & = {\alpha \over 1-\alpha} \, {1-p \over p -q} \, , \qquad
\langle\widehat{\tau}_x\rangle = (1-\alpha) \langle\tau_x\rangle + \alpha\, ,
&\eqnlabel{resDens1D2Dspecb} \cr
}$$
\endeqnseries
for case I or II, respectively.
\mn
Next we discuss the current. The current can be computed in the
interior of the system at those places where hopping processes
are possible during the next time step and is given by
expectation values of the operator
$$\Jop = p \opBh \opA - q \opAh \opB \, .
\label{currentOp}$$
During the first time step it can also be computed at the boundaries
where it is given by expectation values of the operators
$\Jop_L = \alpha \opA - \gamma \opB$ and $\Jop_R = \beta \opBh - \delta \opAh$
respectively.
\mn
First, we use this to compute the current from
the scalar product state \ref{repOneD}.
The result obtained directly from \ref{repOneD} can be written
on the manifold \ref{RelOneD} either as
\begineqnseries[currentOneD]
$$J = \, \beta \, {\frac {(p  - q)  - (\beta+\delta) (1-q +
\kappat_{-}(\beta,\delta))}{\left
(p-q\right )\left (1-\beta-\delta\right )}} \, ,
\label{currentOneDa}$$
or as
$$J = \, \alpha\, {\frac {(p  - q)  - (\alpha+\gamma) (1-q
+ \kappat_{-}(\alpha,\gamma) )}{\left
(p-q\right )\left (1-\alpha-\gamma\right )}} \, .
\label{currentOneDb}$$
\endeqnseries
It is easy to specialize this result to $\gamma=\delta=0$ after
recalling that $\kappat_{-}(x,0) = 0$. In this special case
e.g.\ \ref{currentOneDa} turns into
$J = \, \beta \, {\frac {(p  - q)  - \beta (1-q)}
{\left (p-q\right )\left (1-\beta\right )}}$.
\mn
Analogously the current can be computed from the two-dimensional
representation. In order to obtain the result for the thermodynamic
limit one now has to compute it by taking the matrix
elements $J_L = \left(\alpha \astate{W} \opA - \gamma \astate{W} \opB\right)_{i}/
\left( \astate{W} C\right)_{i}$ at the left boundary, 
$J_R = \left(\beta \opBh \state{V} - \delta \opAh \state{V}\right)_{i}/
\left(C \state{V}\right)_{i}$ at the right boundary and
$J = \left(p \opBh \opA-q \opAh \opB\right)_{i,i}/C_{i,i}^2$
in the interior where $i$ is chosen such that $C_{i,i}$ is
maximal. Inserting \ref{bulkRep2D}, \ref{paramAval} and
\ref{boundVecTwoD} here, one finds the simplest result
at the left boundary for $i=1$.  The current is equal to
\ref{currentOneDa} in case I and to \ref{currentOneDb} in case II
on the manifold where \ref{RelTwoD} is valid.
In the limit of small probabilities this result goes
into the one obtained for sequential dynamics, namely
eqs.\ (4.10) and (4.11) of \cite{\Sandow}. It is not
a coincidence that we have obtained the same expressions
for the current from the one- and two-dimensional representations,
because (as we will show in the Section 6) they are
valid throughout the corresponding phases.
\mn
Finally, we turn to the correlation length $\zeta$ for
the two-dimensional representation which is related to
the eigenvalues of $C$ via $\exp(1/\zeta) = C_{2,2}/C_{1,1}$.
It is given by
$$\exp\left({1 \over \zeta}\right) =
{\kappat_{+}(\alpha,\gamma) \left(1 - p + \kappat_{+}(\beta,\delta)\right)
                            \left(1 - q + \kappat_{+}(\beta,\delta)\right)
\over
\kappat_{+}(\beta,\delta) \left(1 - p + \kappat_{+}(\alpha,\gamma)\right)
                          \left(1 - q + \kappat_{+}(\alpha,\gamma)\right)}
\label{resCorrLen}$$
on the manifold \ref{RelTwoD}. To be precise, we have been able
to establish the validity of \ref{resCorrLen} analytically
for either $\gamma = 0$ or $\delta = 0$. For both rates non-zero
we have performed a careful numerical verification. As expected,
the result of \cite{\EssRi} (eq.\ (102) loc.\ cit.) is recovered in
the limit of small probabilities.
Whether \ref{resCorrLen} is also valid off the manifold specified
by \ref{RelTwoD}, and if so, in which areas, is beyond the scope
of this paper.
\mn
The result \ref{resCorrLen} can also be used to translate the
inequalities between $C_{2,2}$ and $C_{1,1}$ into inequalities
between $\kappat_{+}(\alpha,\gamma)$ and $\kappat_{+}(\beta,\delta)$.
On the manifold \ref{RelTwoD} one finds numerically that
$\kappat_{+}(\alpha,\gamma) \kappat_{+}(\beta,\delta) > (1-p) (1-q)$
\footnote{${}^{2})$}{According to \cite{\EssRi} one expects that
$\kappat_{+}(\alpha,\gamma) \kappat_{+}(\beta,\delta) \ge (1-p) (1-q)$
holds for any finite-dimensional representation whereof infinitely
many should exist. This inequality can be obtained by applying
the argument of the next Section to the corresponding
inequality of \cite{\EssRi}.}.
This can be used to infer from \ref{resCorrLen} that
$C_{2,2} > C_{1,1}$ (case I) iff $\kappat_{+}(\beta,\delta) >
\kappat_{+}(\alpha,\gamma)$ and vice versa.
\bn
\section{Mapping onto the Hamiltonian limit}
\mn
In this Section we will impose a condition similar to \ref{matDiffId}
for general values of the parameters and use this to map the
problem for parallel dynamics to the one with sequential dynamics.
\mn
After fixing the normalization of the representation of the operators
$\opA$, $\opB$, $\opAh$ and $\opBh$ suitably (but different from
the one used in Section 3), we will from now on concentrate on
representations where
$$\opBh - \opB = \opA - \opAh = \id
\label{chooseDiffId}$$
holds (this condition has also been proposed in \cite{\RSS}).
This choice may be a restriction in the sense that
representations of the algebra \ref{parAlg}
might exist that are not equivalent to one where
\ref{chooseDiffId} holds. We will now show that \ref{chooseDiffId}
can be used to lift the Fock space representation of the algebra
\ref{seqAlg} to a representation of the algebra \ref{parAlg} and therefore
one can (at least in principle) compute stationary expectation
values for arbitrary probabilities $\alpha$, $\beta$,
$\gamma$, $\delta$, $p$ and $q$. Even if other representations
of the parallel algebra should exist, they would have to
give rise to the same groundstate that can also be
obtained from a representation where \ref{chooseDiffId}
is valid. Thus, from a physical point of view it is completely
sufficient to study only representations satisfying
\ref{chooseDiffId}.
\mn
We attempt a mapping to the sequential algebra \ref{seqAlg}
by identifying $E$ with a suitable linear combination
of $\opA$ and $\opAh$ and $D$ with another suitable linear
combination of $\opB$ and $\opBh$. This leads to the
following ansatz for the operators $\opA$, $\opAh$,
$\opB$ and $\opBh$ in terms of $E$ and $D$: 
$$\eqalign{
\opA &= n_E E + e \id \, , \qquad
\opAh =  n_E E - (1-e) \id \, , \cr
\opB &= n_D D - d \id \, , \qquad
\opBh = n_D D + (1-d) \id \, , \cr
}\label{mapSeq}$$
where the free constants $n_E$ and $n_D$ reflect the
freedom of choice of relative normalization of the
representations and the constants $e$ and $d$ correspond
to the points of identification. Inserting this ansatz
into \ref{parAlgI} one obtains precisely one independent
relation between $E$ and $D$:
$$\eqalign{
n_E n_D (p D E - q E D) =&
    n_E E \left( d (p-q) + (1-p) \right)
 + n_D D \left(-e (p-q) + (1-q) \right) \cr
& + e d (p-q) + e (1-p) - d (1-q) \, . \cr
}\label{insMap}$$
In order to be able to identify this with \ref{seqAlg}
the constant term in \ref{insMap} must vanish.
This is ensured by chosing
$$e = {d (1-q) \over d (p-q) + (1-p)} \, .
\label{MapConVan}$$
Also the linear term in $E$ must appear with the
same coefficient as the linear term in $D$
on the r.h.s.\ of \ref{insMap}. This is the case if
$$n_D = -n_E {d (p-q) + (1-p) \over e (p-q) - (1-q)}
   = n_E {\left(d (p-q) + (1-p) \right)^2 \over
          (1-p) (1-q)} \, .
\label{MapLinVan}$$
Using \ref{MapConVan} and \ref{MapLinVan} the relation
\ref{insMap} now reads
$$n_E {d (p-q) + (1-p) \over (1-p) (1-q)}
   \left( p D E - q E D \right) = E + D \, .
\label{renMap}$$
Inserting the ansatz \ref{mapSeq} into the boundary equations
\ref{parAlgB} leads to
$$\astate{W} {\alpha n_E E - \gamma n_D D \over 1 - \alpha e - \gamma d}
= \astate{W} \, , \qquad
{\beta n_D D - \delta n_E E \over 1 + \delta e + \beta d - (\delta + \beta)}
\state{V} = \state{V} \, .
\label{insBound}$$
The equations \ref{renMap}
and \ref{insBound} are identical to the algebra of the
sequential limit \ref{seqAlg} if one identifies the rates
in that limit as follows
$$\eqalign{
\ph &= n_E {d (p-q)+(1-p) \over (1-p) (1-q)} p \, , \qquad
\qh = n_E {d (p-q)+(1-p) \over (1-p) (1-q)} q \, , \cr
\alphah &= n_E {1 \over 1-\alpha e-\gamma d} \alpha \, , \qquad\qquad
\betah = n_D {1 \over 1+\delta e+\beta d-(\delta+\beta)} \beta \, , \cr
\gammah &= n_D {1 \over 1-\alpha e-\gamma d} \gamma \, , \qquad\qquad
\deltah = n_E {1 \over 1+\delta e+\beta d-(\delta+\beta)} \delta \, . \cr
}\label{renRates}$$
Here, the constants $e$ and $n_D$ are fixed by \ref{MapConVan} and
\ref{MapLinVan}. The constant $n_E$ remains free and reflects the
freedom of normalization of the algebra in the sequential limit.
It is possible to choose $n_E$ such that the `renormalized' rates
in the bulk are equal to the hopping probabilities, i.e.\
$\ph = p$ and $\qh = q$. If one does not fix $n_E$ in a computation
of a physical quantity, it has to disappear from the final
result. Also the constant $d$ is still free in \ref{renRates}, but
for explicit computations it will, in contrast to $n_E$, be
fixed to a convenient value that ensures $n_E = n_D$. Then, the
operators $C$ for the parallel and sequential case may be identified
with each other.
\mn
From \ref{renRates} one sees that in the sequential limit the
renormalized parameters differ from the initial probabilities only
by a factor $n_E$, which also equals $n_D$ because of \ref{MapLinVan}.
Therefore, choosing $n_E = 1$, both sets of parameters and also
the matrix-product states become identical.
\mn
One can also check that the conditions of Section 3 for having
a one- or two-dimensional representation can be recovered from
known results. All one has to do is insert \ref{renRates}
where $n_E$ and $d$ are kept as free parameters into the
conditions for the sequential limit \cite{\EssRi}. The $d$-independent
factors of these renormalized conditions are precisely \ref{RelOneD}
and \ref{RelTwoD} respectively. So, the simple scalar product
state as well as the two-dimensional matrix-product state of
the general case can be obtained by a simple `renormalization'
process from the ones of the sequential limit. The same holds
also for higher-dimensional representations whose existence was
shown in \cite{\EssRi}.
\bn
\section{Current and phase diagram}
\mn
Now we show how to use the mapping of the previous Section to compute
quantities of physical interest.
\mn
According to \cite{\Sandow,\EssRi}, the phase diagram in the
sequential case is given by a single function $\kappa_{+}$ of
the rates at either boundary, i.e.\ by
$\kappa_{+}(\alphah,\gammah)$ and $\kappa_{+}(\betah,\deltah)$.
The same will hold in  parallel case, if the quantities \ref{renRates}
are inserted into the functions $\kappa_{+}$. This statement has already
been partially verified in Section 4 where the variables
$\kappat_{\pm}(x,y)$ have been used. To be more precise, the parallel
case will be descibed by functions
$\kappah_{\pm}(\alpha,\gamma) = \kappa_{\pm}(\alphah,\gammah)$
and $\kappah_{\pm}(\beta,\delta) = \kappa_{\pm}(\betah,\deltah)$ 
which are obtained from the $\kappa_{\pm}$ of the sequential
case \cite{\Sandow,\EssRi} by choosing the constant $d$ in the
mapping \ref{renRates} such that $n_D = n_E$. The function
$\kappah_{\pm}$ is given by
$$\kappah_{\pm}(x,y) = {\kappat_{\pm}(x,y) \over
    \sqrt{\left (1-q\right )\left (1-p\right )}} \, ,
\label{defKappaH}$$
where $\kappat_{\pm}(x,y)$ was defined in \ref{defKappaT}.
Recall from the beginning of Section 4 that $\kappah_{+}(\alpha,\gamma)$
and $\kappah_{+}(\beta,\delta)$ may be regarded as effective
input/output rates at the boundaries.
Recall also that the $\kappat_{\pm}$ and $\kappah_{\pm}$
should be used only for $p > q$ and that the results for
the case $p < q$ can be obtained by applying parity.
\mn
It is straightforward to compute the currents for the general
case with parallel dynamics using the results of \cite{\Sandow}
for the sequential case. The operators needed for that
have already been given in \ref{currentOp}
for the bulk and for the boundaries below that equation.
Using \ref{parAlgB} and the ansatz \ref{chooseDiffId}
one can check that 
$$\eqalign{
J_L
&= {\astate{W} \left(\alpha \opA - \gamma \opB\right) C^{N-1} \state{V}
    \over \astate{W} C^N \state{V} }
 = {\astate{W} C^{N-1} \state{V} \over \astate{W} C^N \state{V} } = J \cr
&= {\astate{W} C^{N-1} \left(\beta \opBh - \delta \opAh \right) \state{V}
    \over \astate{W} C^N \state{V} } = J_R \, . \cr
}\label{currentFormula}$$
This means that the current is simply given by ratios of the
normalization constants $\astate{W} C^N \state{V}$ for different $N$
if the condition \ref{chooseDiffId} holds. Thus, in order to
compute the current in the thermodynamic limit, the behaviour
of such ratios has to be studied for large $N$.
\mn
One finds from \ref{mapSeq} that
$$C = \opA + \opB = n_E (E+D) + (e-d) \, .
\label{CseqPar}$$
In \cite{\Sandow} the operators $D$ and $E$ have been expressed
in terms of the creation and annihilation operators $F$ and $\FD$
of a $q$-deformed harmonic oscillator by setting
$$D = {F + 1 \over \ph - \qh} \, , \qquad
E = {\FD + 1 \over \ph - \qh} \, .
\label{insqHarm}$$
Inserting this into \ref{CseqPar} leads to
$$C = n_E {F + \FD + \lambda \over \ph - \qh}
\label{CqHarm}$$
with
$$\lambda = 2 + (e-d) {\ph - \qh \over n_E}
  = {2 - q - p \over \sqrt{(1-q)(1-p)}} \, .
\label{lambdaVal}$$
First we consider the case $\kappah_{+}(\alpha,\gamma) < 1$
and $\kappah_{+}(\beta,\delta) < 1$.
According to \cite{\Sandow} this is the maximal current phase
and to obtain $J$, the same computation as there can be used.
In eq.\ (3.39) loc.\ cit.\ the coefficient
$2$ of $c_{ik}^L$ has to be replaced by $\lambda$ because
of the extra constant in \ref{CqHarm}. Tracing the effect
of this modification one finds that for parallel dynamics the
current in the maximal current phase is given by
$$J = {1 \over n_E} \; {\ph - \qh \over 2 + \lambda}
= {\sqrt{1-q} - \sqrt{1-p} \over \sqrt{1-q} + \sqrt{1-p}} \, .
\label{maxCurrent}$$
For small $p$ and $q$ the result (4.6) of \cite{\Sandow} is recovered.
\mn
This result can be checked \cite{\KoelnB} using the one-dimensional
representation because the line \ref{RelOneD} touches the region
where \ref{maxCurrent} is valid (see Fig.\ 3 below).
\mn
The currents for the other two phases can be obtained from
similar modifications of the computations in \cite{\Sandow}.
First we consider the case
$\kappah_{+}(\beta,\delta) > \kappah_{+}(\alpha,\gamma)$
and $\kappah_{+}(\beta,\delta) > 1$.
The relation (4.7) of \cite{\Sandow} remains valid with the
new parameters and reads
$$\astate{W} C^{N-1} F \state{V} \approx \kappa_{+}(\betah,\deltah)
  \astate{W} C^{N-1}\state{V}
  = \kappah_{+}(\beta,\delta) \astate{W} C^{N-1}\state{V} \, .
\label{ren4.3Sandow}$$
One can further rewrite $C$ as follows using \ref{CseqPar}
-- \ref{lambdaVal}
$$\eqalign{
C =& {n_E \over \deltah} \left(
- (\betah D - \deltah E) + {\betah + \deltah  +
  \deltah \, (\ph - \qh) (e-d) / n_E \over \ph - \qh}
+ {\betah + \deltah \over \ph - \qh} F \right) \cr
=& {n_E \over \deltah} \left(
- (\betah D - \deltah E) + {\betah + \deltah
  + \deltah (\lambda - 2) \over \ph - \qh}
+ {\betah + \deltah \over \ph - \qh} F \right) \, . \cr
}\label{Crewrite}$$
Making first use of \ref{Crewrite} and then of
\ref{ren4.3Sandow} one finds the following recurrence
relation for the normalization constants (generalizing
eqs.\ (4.8) and (4.9) of \cite{\Sandow})
$$\eqalign{
\astate{W} & C^N \state{V} \cr
= &
{n_E \over \deltah} \; {1 \over \ph - \qh} \left\{\left(
\betah + \deltah+ \deltah (\lambda - 2) - \ph + \qh \right)
\astate{W} C^{N-1}\state{V} + \left(\betah + \deltah\right)
\astate{W} C^{N-1} F \state{V} \right\} \cr
\approx &
{n_E \over \deltah} \; {1 \over \ph - \qh} \left\{\left(
\betah + \deltah+ \deltah (\lambda - 2) - \ph + \qh \right)
+ \left(\betah + \deltah\right) \kappah_{+}(\beta,\delta)
\right\} \astate{W} C^{N-1} \state{V} \, .
}\label{CrecurrP}$$
Inserting this result into \ref{currentFormula} one
recovers after a straightforward computation the result
\ref{currentOneDa} that we already obtained from the
one- and two-dimensional representations.
The result for the other phase $\kappah_{+}(\alpha,\gamma) >
\kappah_{+}(\beta,\delta)$ and $\kappah_{+}(\alpha,\gamma) > 1$
can be obtained simply by replacing $\beta$ by $\alpha$ and $\delta$
by $\gamma$ and coincides with \ref{currentOneDb}.
\mn
These results can be used to draw the phase diagram which turns
out to be essentially the same as the one for the sequential
limit \cite{\Sandow,\EssRi}. It is shown in Fig.\ 3
in terms of the variables $\kappah_{+}(x,y)$, using a logarithmic
scale. There are at least
three distinct phases which are characterized only by the values of
$\kappah_{+}(\alpha,\gamma)$ and $\kappah_{+}(\beta,\delta)$:
\sn
\item{I:} \qquad
$\kappah_{+}(\beta,\delta) > \kappah_{+}(\alpha,\gamma)$,
$\kappah_{+}(\beta,\delta) > 1$ (high density)
\item{II:} \qquad
$\kappah_{+}(\alpha,\gamma) > \kappah_{+}(\beta,\delta)$,
$\kappah_{+}(\alpha,\gamma) > 1$ (low density)
\item{III:} \qquad
$\kappah_{+}(\alpha,\gamma) < 1$,
$\kappah_{+}(\beta,\delta) < 1$ (maximal current)
\sn
The density in phases I and II is approximately constant in
the bulk. On the coexistence line $\kappah_{+}(\alpha,\gamma) =
\kappah_{+}(\beta,\delta)$ which separates these two phases
the density profile is linear in space. 
\mn
Phases I and II can be mapped onto each other using parity
and particle-hole symmetry. This `duality' transformation exchanges
$\tau_x \leftrightarrow 1 - \widehat{\tau}_{N+1-x}$,
$\alpha \leftrightarrow \beta$, $\gamma \leftrightarrow \delta$
and keeps $p$ and $q$ unchanged. In particular
$\kappah_{+}(\alpha,\gamma)$ is exchanged with $\kappah_{+}(\beta,\delta)$
from which one obtains the mapping between the two phases.
\mn
The dotted line in Fig.\ 3 shows the condition \ref{RelOneD}
for the existence of a one-dimensional representation which
can be cast in the form $\kappah_{+}(\alpha,\gamma)
\kappah_{+}(\beta,\delta) = 1$. The figure also shows the
condition \ref{RelTwoD} for having a two-dimensional representation
with two choices of $p$ and $q$. In both cases
$\kappah_{+}(\alpha,\gamma) \kappah_{+}(\beta,\delta) = p/q$
is verified numerically. For $\gamma=\delta=0$, it can also be
shown analytically that this condition is equivalent to
\ref{RelTwoD}. This suggests that the condition for having
a two-dimensional representation is a hyperbola in the
$\kappah_{+}(\alpha,\gamma)$-$\kappah_{+}(\beta,\delta)$
plane. By changing the value of $p/q$ this hyperbola can
be swept over the entire region above the dotted
line marking the one-dimensional representation.
\mn
\mn
$$
\Horline{0}{0}{70}
\Horline{0}{70}{70}
\Verline{0}{0}{70}
\Verline{70}{0}{70}
\verline{35}{0}{35}
\horline{0}{35}{35}
\putbox{17.5}{17.5}{{\bigf III}}
\putbox{30}{55}{{\bigf I}}
\putbox{55}{30}{{\bigf II}}
\putbox{35}{-8}{$\kappah_{+}(\alpha,\gamma)$}
\putbox{17.5}{-3}{$10^{-1}$}
\putbox{35}{-3}{$1^{\phantom{1}}$}
\putbox{52.5}{-3}{$10^{\phantom{1}}$}
\putbox{70}{-3}{$10^{2}$}
\Verline{17.5}{0}{2}
\Verline{35}{0}{2}
\Verline{52.5}{0}{2}
\Verline{17.5}{68}{2}
\Verline{35}{68}{2}
\Verline{52.5}{68}{2}
\rightputbox{-9}{35}{$\kappah_{+}(\beta,\delta)$}
\rightputbox{-1.5}{-3}{$10^{-2}$}
\rightputbox{-1.5}{17.5}{$10^{-1}$}
\rightputbox{-1.5}{35}{$1^{\phantom{1}}$}
\rightputbox{-1.5}{52.5}{$10^{\phantom{1}}$}
\rightputbox{-1.5}{70}{$10^{2}$}
\Horline{0}{17.5}{2}
\Horline{0}{35}{2}
\Horline{0}{52.5}{2}
\Horline{68}{17.5}{2}
\Horline{68}{35}{2}
\Horline{68}{52.5}{2}
\dline{36.75}{36.75}
\dline{40.25}{40.25}
\dline{43.75}{43.75}
\dline{47.25}{47.25}
\dline{50.75}{50.75}
\dline{54.25}{54.25}
\dline{57.75}{57.75}
\dline{61.25}{61.25}
\dline{64.75}{64.75}
\dline{68.25}{68.25}
\plot{1}{69}
\plot{2}{68}
\plot{3}{67}
\plot{4}{66}
\plot{5}{65}
\plot{6}{64}
\plot{7}{63}
\plot{8}{62}
\plot{9}{61}
\plot{10}{60}
\plot{11}{59}
\plot{12}{58}
\plot{13}{57}
\plot{14}{56}
\plot{15}{55}
\plot{16}{54}
\plot{17}{53}
\plot{18}{52}
\plot{19}{51}
\plot{20}{50}
\plot{21}{49}
\plot{22}{48}
\plot{23}{47}
\plot{24}{46}
\plot{25}{45}
\plot{26}{44}
\plot{27}{43}
\plot{28}{42}
\plot{29}{41}
\plot{30}{40}
\plot{31}{39}
\plot{32}{38}
\plot{33}{37}
\plot{34}{36}
\plot{35}{35}
\plot{36}{34}
\plot{37}{33}
\plot{38}{32}
\plot{39}{31}
\plot{40}{30}
\plot{41}{29}
\plot{42}{28}
\plot{43}{27}
\plot{44}{26}
\plot{45}{25}
\plot{46}{24}
\plot{47}{23}
\plot{48}{22}
\plot{49}{21}
\plot{50}{20}
\plot{51}{19}
\plot{52}{18}
\plot{53}{17}
\plot{54}{16}
\plot{55}{15}
\plot{56}{14}
\plot{57}{13}
\plot{58}{12}
\plot{59}{11}
\plot{60}{10}
\plot{61}{9}
\plot{62}{8}
\plot{63}{7}
\plot{64}{6}
\plot{65}{5}
\plot{66}{4}
\plot{67}{3}
\plot{68}{2}
\plot{69}{1}
\diamond{15.7500000000000}{62.5996219575942}
\diamond{62.5996219575942}{15.7500000000000}
\diamond{17.5000000000000}{60.8496219575944}
\diamond{60.8496219575944}{17.5000000000000}
\diamond{19.2500000000000}{59.0996219575942}
\diamond{59.0996219575942}{19.2500000000000}
\diamond{21.0000000000000}{57.3496219575944}
\diamond{57.3496219575944}{21.0000000000000}
\diamond{22.7500000000000}{55.5996219575948}
\diamond{55.5996219575948}{22.7500000000000}
\diamond{24.5000000000000}{53.8496219575946}
\diamond{53.8496219575946}{24.5000000000000}
\diamond{26.2500000000000}{52.0996219575944}
\diamond{52.0996219575944}{26.2500000000000}
\diamond{28.0000000000000}{50.3496219575943}
\diamond{50.3496219575943}{28.0000000000000}
\diamond{29.7500000000000}{48.5996219575943}
\diamond{48.5996219575943}{29.7500000000000}
\diamond{31.5000000000000}{46.8496219575943}
\diamond{46.8496219575943}{31.5000000000000}
\diamond{33.2500000000000}{45.0996219575943}
\diamond{45.0996219575943}{33.2500000000000}
\diamond{35.0000000000000}{43.3496219575946}
\diamond{43.3496219575946}{35.0000000000000}
\diamond{36.7500000000000}{41.5996219575944}
\diamond{41.5996219575944}{36.7500000000000}
\diamond{38.5000000000000}{39.8496219575944}
\diamond{39.8496219575944}{38.5000000000000}
\square{8.7500000000000}{64.3315970334749}
\square{64.3315970334749}{8.7500000000000}
\square{10.5000000000000}{62.5815970334748}
\square{62.5815970334748}{10.5000000000000}
\square{12.2500000000000}{60.8315970334749}
\square{60.8315970334749}{12.2500000000000}
\square{14.0000000000000}{59.0815970334748}
\square{59.0815970334748}{14.0000000000000}
\square{15.7500000000000}{57.3315970334751}
\square{57.3315970334751}{15.7500000000000}
\square{17.5000000000000}{55.5815970334749}
\square{55.5815970334749}{17.5000000000000}
\square{19.2500000000000}{53.8315970334748}
\square{53.8315970334748}{19.2500000000000}
\square{21.0000000000000}{52.0815970334748}
\square{52.0815970334748}{21.0000000000000}
\square{22.7500000000000}{50.3315970334746}
\square{50.3315970334746}{22.7500000000000}
\square{24.5000000000000}{48.5815970334748}
\square{48.5815970334748}{24.5000000000000}
\square{26.2500000000000}{46.8315970334746}
\square{46.8315970334746}{26.2500000000000}
\square{28.0000000000000}{45.0815970334747}
\square{45.0815970334747}{28.0000000000000}
\square{29.7500000000000}{43.3315970334746}
\square{43.3315970334746}{29.7500000000000}
\square{31.5000000000000}{41.5815970342415}
\square{41.5815970342415}{31.5000000000000}
\square{33.2500000000000}{39.8315970334747}
\square{39.8315970334747}{33.2500000000000}
\square{35.0000000000000}{38.0815970334749}
\square{38.0815970334749}{35.0000000000000}
\square{36.7500000000000}{36.3315970334748}
\hskip70mm
$$
\sn

{\par\noindent\figindents
{\bf Fig.\ 3:} Phase diagram of the kinetic model.
The dotted line shows the condition \ref{RelOneD} for
a one-dimensional representation. The condition \ref{RelTwoD} for
a two-dimensional representation is shown for two choices of
$p$ and $q$: The symbol `\SymbolB' is for $p=0.75$, $q=0.25$
and the symbol `\SymbolA' is for $p=0.6$, $q=0.4$.
\par\noindent}
\mn
For a more detailed discussion of the
case $\gamma=\delta=0$ we refer to \cite{\KoelnB}.
\mn
Our results for the current and densities are summarized in Table 1.
The correlation length is given by \ref{resCorrLen}, but probably
not throughout the entire phases I and II. Presumably one has to
choose at least $\kappah_{+}(\alpha,\gamma) \kappah_{+}(\beta,\delta) > 1$
because on the line
$\kappah_{+}(\alpha,\gamma) \kappah_{+}(\beta,\delta) = 1$ there is
a scalar product state and the correlation length diverges as one
approaches this line.
\mn
All these results have the correct behaviour under the duality
transformation which exchanges the phases I and II.
\mn
\centerline{\vbox{
\hbox{
\vrule \hskip 1pt
\vbox{ \offinterlineskip
\def\tablespace{height2pt&\omit&&\omit&&\omit&&\omit&\cr}
\def\tablerule{ \tablespace
                \noalign{\hrule}
                \tablespace        }
\hrule
\halign{&\vrule#&
  \strut\hskip 4pt\hfil#\hfil\hskip 4pt\cr
\tablespace
& {\it Phase} && $J$ && $\langle\tau_x\rangle$ && $\langle\widehat{\tau}_x\rangle$ 
                                          & \cr \tablespace \tablerule
& I     && $\displaystyle \beta \, {\frac {(p  - q)  - (\beta+\delta) (1-q +
 \kappat_{-}(\beta,\delta))}{\left (p-q\right )\left (1-\beta-\delta\right )}}$
  && $\displaystyle {\kappat_{+}(\beta,\delta) \over 1 - q + \kappat_{+}(\beta,\delta)}$
    && $\displaystyle {\kappat_{+}(\beta,\delta) \over 1 - p + \kappat_{+}(\beta,\delta)}$
      & \cr\tablespace\tablespace
& II    && $\displaystyle \alpha\, {\frac {(p  - q)  - (\alpha+\gamma) (1-q +
 \kappat_{-}(\alpha,\gamma) )}{\left (p-q\right )\left (1-\alpha-\gamma\right )}}$
  && $\displaystyle {1-p \over 1 - p + \kappat_{+}(\alpha,\gamma)}$
    && $\displaystyle {1-q \over 1 - q + \kappat_{+}(\alpha,\gamma)}$
      & \cr\tablespace\tablespace
& III   && $\displaystyle {\sqrt{1-q} - \sqrt{1-p} \over \sqrt{1-q} + \sqrt{1-p}}$
  && \hbox to 1.5 cm{\hrulefill} && \hbox to 1.5 cm{\hrulefill} & \cr\tablespace
}
\hrule}\hskip 1pt \vrule}
\hbox{\quad {\bf Table 1:} Summary of our results for the current $J$,
                    the bulk density $\langle\tau_x\rangle$}
\hbox{\quad \phantom{{\bf Table 1:}} on the odd sublattice and the
                    bulk density $\langle\widehat{\tau}_x\rangle$ on the even}
\hbox{\quad \phantom{{\bf Table 1:}} sublattice in the three phases.}}
}
\mn
We mention that the result for the current can also be obtained
by using the bulk-densities in the mean-field formula
$J = p \langle\widehat{\tau}_x\rangle (1-\langle\tau_{x+1}\rangle)
   - q (1-\langle\widehat{\tau}_x\rangle) \langle\tau_{x+1}\rangle$.
In regions I and II this can be shown analytically for
$\gamma = \delta = 0$. In region III it can be verified numerically.
\mn
In appendix B we show how to use \ref{matDiffId} with $g = 1$
to compute correlation functions for the case of symmetric
diffusion $p=q$ where the results of this Section and Section 4
may not be applied directly (compare also the corresponding
remark in Section 3). For this special case, the algebra is
much simpler.
Appendix C describes how one can use the same representation to compute
physical quantities on finite lattices with system sizes up to
a few hundred sites. This can be used to check the validity
of our results in Table 1 as well as the correlation length
\ref{resCorrLen} for
$\kappah_{+}(\alpha,\gamma) \kappah_{+}(\beta,\delta) > 1$
numerically.
\bn
\section{Discussion}
\mn
We have considered the diffusion of hard-core particles between
two reservoirs for a particular kind of parallel dynamics.
The results show that the stationary state has a similar
matrix-product form as in the sequential case and can actually
be obtained from that limit. Therefore the physical properties,
in particular the phase diagram, are also similar although the
formulae are more involved. 
\mn
In \cite{\RSS} it was found that the same is true for yet
another type of dynamics, where the stochastic motion takes
place step-by-step along the chain. This can be visualized
nicely in the vertex-model picture as shown in Fig.\ 4. The
processes take place in a diagonal strip of the lattice
and the time-evolution operator is seen to be the usual
row-to-row transfer matrix $T_{\rm row}$ of the vertex model,
with a shift in the numbering of the upper row of variables
and additional boundary vertices. Using the same exchange
mechanism at each vertex as in Section 2, one finds that
the pair of operators $\opA$ and $\opB$ only appears in the
intermediate steps and the matrix-product state becomes
a homogenous state involving only the operators $\opAh$
and $\opBh$. An independent treatment of this matrix-product
state would involve computations similar to those that we
have presented here. Alternatively, one can also directly
use our results for the current and correlation length for
the updates as in Fig.\ 4 (the densities are only the same
on the even sublattice) \cite{\RSS}.
\mn
\mn
$$
\lhvertex{20}{3}
\vertex{34}{3}
\vertex{48}{3}
\Brhvertex{62}{3}
\vertex{27}{10}
\vertex{41}{10}
\Bvertex{55}{10}
\lhvertex{20}{17}
\vertex{34}{17}
\Bvertex{48}{17}
\rhvertex{62}{17}
\vertex{27}{24}
\Bvertex{41}{24}
\vertex{55}{24}
\lhvertex{20}{31}
\Bvertex{34}{31}
\vertex{48}{31}
\rhvertex{62}{31}
\Bvertex{27}{38}
\vertex{41}{38}
\vertex{55}{38}
\Blhvertex{20}{45}
\vertex{34}{45}
\vertex{48}{45}
\rhvertex{62}{45}
\putbox{23.5}{32.5}{$\scriptstyle 1$}
\putbox{30.5}{25.5}{$\scriptstyle 2$}
\putbox{37.5}{18.5}{$\scriptstyle 3$}
\plot{47}{9}
\plot{48}{8}
\plot{49}{7}
\putbox{58.5}{-2.5}{$\scriptstyle N$}
\putbox{23.5}{50.5}{$\scriptstyle 1$}
\putbox{30.5}{43.5}{$\scriptstyle 2$}
\putbox{37.5}{36.5}{$\scriptstyle 3$}
\plot{47}{27}
\plot{48}{26}
\plot{49}{25}
\putbox{58.5}{15.5}{$\scriptstyle N$}
\dashdiaglineSevenFive{62}{-4}
\dashdiaglineSevenFive{61.5}{-3.5}
\dashdiaglineSevenFive{61}{-3}
\dashdiaglineSevenFive{60.5}{-2.5}
\dashdiaglineSevenFive{60}{-2}
\dashdiaglineSevenFive{59.5}{-1.5}
\dashdiaglineSevenFive{59}{-1}
\dashdiaglineSevenFive{58.5}{-0.5}
\dashdiaglineSevenFive{58}{0}
\dashdiaglineSevenFive{57.5}{0.5}
\dashdiaglineSevenFive{57}{1}
\dashdiaglineSevenFive{56.5}{1.5}
\dashdiaglineSevenFive{56}{2}
\dashdiaglineSevenFive{55.5}{2.5}
\dashdiaglineSevenFive{55}{3}
\dashdiaglineSevenFive{54.5}{3.5}
\dashdiaglineSevenFive{54}{4}
\dashdiaglineSevenFive{53.5}{4.5}
\dashdiaglineSevenFive{53}{5}
\dashdiaglineSevenFive{52.5}{5.5}
\dashdiaglineSevenFive{52}{6}
\dashdiaglineSevenFive{51.5}{6.5}
\dashdiaglineSevenFive{51}{7}
\dashdiaglineSevenFive{50.5}{7.5}
\dashdiaglineSevenFive{50}{8}
\dashdiaglineSevenFive{49.5}{8.5}
\dashdiaglineSevenFive{49}{9}
\dashdiaglineSevenFive{48.5}{9.5}
\dashdiaglineSevenFive{48}{10}
\dashdiaglineSevenFive{47.5}{10.5}
\dashdiaglineSevenFive{47}{11}
\dashdiaglineSevenFive{46.5}{11.5}
\dashdiaglineSevenFive{46}{12}
\dashdiaglineSevenFive{45.5}{12.5}
\dashdiaglineSevenFive{45}{13}
\dashdiaglineSevenFive{44.5}{13.5}
\dashdiaglineSevenFive{44}{14}
\dashdiaglineSevenFive{43.5}{14.5}
\dashdiaglineSevenFive{43}{15}
\dashdiaglineSevenFive{42.5}{15.5}
\dashdiaglineSevenFive{42}{16}
\dashdiaglineSevenFive{41.5}{16.5}
\dashdiaglineSevenFive{41}{17}
\dashdiaglineSevenFive{40.5}{17.5}
\dashdiaglineSevenFive{40}{18}
\dashdiaglineSevenFive{39.5}{18.5}
\dashdiaglineSevenFive{39}{19}
\dashdiaglineSevenFive{38.5}{19.5}
\dashdiaglineSevenFive{38}{20}
\dashdiaglineSevenFive{37.5}{20.5}
\dashdiaglineSevenFive{37}{21}
\dashdiaglineSevenFive{36.5}{21.5}
\dashdiaglineSevenFive{36}{22}
\dashdiaglineSevenFive{35.5}{22.5}
\dashdiaglineSevenFive{35}{23}
\dashdiaglineSevenFive{34.5}{23.5}
\dashdiaglineSevenFive{34}{24}
\dashdiaglineSevenFive{33.5}{24.5}
\dashdiaglineSevenFive{33}{25}
\dashdiaglineSevenFive{32.5}{25.5}
\dashdiaglineSevenFive{32}{26}
\dashdiaglineSevenFive{31.5}{26.5}
\dashdiaglineSevenFive{31}{27}
\dashdiaglineSevenFive{30.5}{27.5}
\dashdiaglineSevenFive{30}{28}
\dashdiaglineSevenFive{29.5}{28.5}
\dashdiaglineSevenFive{29}{29}
\dashdiaglineSevenFive{28.5}{29.5}
\dashdiaglineSevenFive{28}{30}
\dashdiaglineSevenFive{27.5}{30.5}
\dashdiaglineSevenFive{27}{31}
\dashdiaglineSevenFive{26.5}{31.5}
\dashdiaglineSevenFive{26}{32}
\dashdiaglineSevenFive{25.5}{32.5}
\dashdiaglineSevenFive{25}{33}
\dashdiaglineSevenFive{24.5}{33.5}
\dashdiaglineSevenFive{24}{34}
\dashdiaglineSevenFive{23.5}{34.5}
\dashdiaglineSevenFive{23}{35}
\dashdiaglineSevenFive{22.5}{35.5}
\dashdiaglineSevenFive{22}{36}
\dashdiaglineSevenFive{21.5}{36.5}
\dashdiaglineSevenFive{21}{37}
\dashdiaglineSevenFive{20.5}{37.5}
\dashdiaglineSevenFive{20}{38}
\dashdiaglineSevenFive{19.5}{38.5}
\dashdiaglineSevenFive{19}{39}
\dashdiaglineSevenFive{18.5}{39.5}
\dashdiaglineSevenFive{18}{40}
\dashdiaglineSevenFive{17.5}{40.5}
\dashdiaglineSevenFive{17}{41}
\dashdiaglineSevenFive{16.5}{41.5}
\dashdiaglineSevenFive{16}{42}
\dashdiaglineSevenFive{15.5}{42.5}
\dashdiaglineSevenFive{15}{43}
\dashdiaglineSevenFive{14.5}{43.5}
\dashdiaglineSevenFive{14}{44}
\dashdiaglineSevenFive{13.5}{44.5}
\dashdiaglineSevenFive{13}{45}
\putbox{76.8}{-2}{\circle\char"21}
\putbox{72.2}{0}{\UArrow}
\putbox{72}{-7}{$T_{\rm row}$}
\hskip 82mm
$$
\sn

{\par\noindent\figindents
{\bf Fig.\ 4:} Representation of a step-by-step dynamics
starting at the right end of the system as a vertex model.
\par\noindent}
\mn
In order to illustrate the connection to vertex models further,
let us also present a graphical representation of the
matrix-product mechanism used here. Denote the pair of operators
$\opA$, $\opB$ by a `\bbullet' and the pair $\opAh$ and $\opBh$
by a `\bcirc'. A dotted
line between them indicates that their product is to
be taken while a termination of the line inside them
means that the vectors $\astate{W}$ and $\state{V}$
have to be multiplied from the left and right respectively.
Then \ref{IntParEq} and \ref{BoundParEq} can be depicted
as follows (compare also Fig.\ 3 of \cite{\Baxter}): 
$$\putbox{10}{6}{\phantom{.}}
\putbox{10}{-6}{\phantom{.}}
\lhvertex{53.5}{0}
\DashhorlineFive{57}{-3.5}
\putbox{57}{-3.5}{\bbullet}
\putbox{64}{0}{$=$}
\DashhorlineFive{68}{0}
\putbox{68}{0}{\bcirc}
\putbox{75}{0}{,}
\DashhorlineFive{85}{-3.5}
\putbox{90}{-3.5}{\bcirc}
\rhvertex{93.5}{0}
\putbox{97}{0}{$=$}
\DashhorlineFive{101}{0}
\putbox{106}{0}{\bbullet}
\putbox{109}{0}{.}
\vertex{10.5}{0}
\DashhorlineFive{3}{-3.5}
\DashhorlineFive{8}{-3.5}
\DashhorlineFive{13}{-3.5}
\putbox{7}{-3.5}{\bcirc}
\putbox{14}{-3.5}{\bbullet}
\putbox{20}{0}{$=$}
\DashhorlineFive{24}{0}
\DashhorlineFive{29}{0}
\DashhorlineFive{34}{0}
\putbox{28}{0}{\bbullet}
\putbox{35}{0}{\bcirc}
\putbox{41}{0}{,}
\hskip 112 mm
\label{GrParEq}$$
The first identity is strikingly similar
to the Yang-Baxter equation (see e.g.\ Chapter 2.3 of
\cite{\Ma}) which for vertex models reads graphically
$$\vertex{18}{0}
\horline{11}{-3.50}{14}
\dline{12.75}{-5.25}
\ddline{23.25}{-5.25}
\putbox{15}{-5}{$u$}
\putbox{21}{-5}{$v$}
\putbox{32}{0}{$=$}
\vertex{46}{0}
\horline{39}{3.5}{14}
\ddline{40.75}{5.25}
\dline{51.25}{5.25}
\putbox{43}{5}{$v$}
\putbox{49}{5}{$u$}
\putbox{53}{0}{.}
\hskip 64 mm
\label{YBE}$$
In both cases one can `pull' a line through the vertex and
the corresponding quantities exchange places. The boundary
Yang-Baxter equations, involving the so-called $K$-matrices,
are more complicated.
This leads us to the question of integrability of the model.
As already mentioned, the time evolution operator
in the sequential limit is a Heisenberg Hamiltonian with
boundary fields. It is therefore integrable in the sense that
it belongs to a whole family of commuting operators
\cite{\GoVe,\InKo}. One would expect that his
also holds for the full vertex model. A proof would have
to follow the lines of \cite{\GoVe-\BaYu}. In any case,
the integrability would be more relevant for the time behaviour
of the system than for the stationary state which we determined.
\mn
Occasionally, it has been conjectured that there is a connection
between the existence of matrix-product states and integrability.
However, examples of scalar product states like in \cite{\PeRy}
show that this cannot be true in general. Nor does the construction
in \cite{\Giac} ensure integrability. It amounts to transforming
the bulk relation in \ref{GrParEq} into a commutator as in \ref{YBE}
by going over from the eigenvector $\state{\Phi}$ to the operator
$P = \state{\Phi}\astate{\Phi}$.
However, this does not help in finding other eigenstates because
$P$ is a simple projector whose unique eigenvector with eigenvalue $1$
is $\state{\Phi}$ while a highly degenerate eigenvalue $0$
accounts for all other vectors.
\mn
Finally, one may ask if the parallel dynamics considered in this
work might be used in other related problems. Here, two situations
come to mind. One would be a model with coagulation and
decoagulation in addition to the hopping processes. With a
proper tuning of the parameters, this is a free-fermion
problem in the sequential limit \cite{\HiKrPe} and the
stationary state has been shown to have a matrix-product
form with four-dimensional matrices \cite{\HiPeSa}.
As in the present work, there are four matrices involved,
two for the homogenous matrix-product state and two in
the generalization of the cancellation mechanism \ref{BoundHamEq}
and \ref{IntHamEq}. Indeed, if the mechanism \ref{BoundParEq}
and \ref{IntParEq} should be applicable to more general
situations than discussed here, one would in general
not expect the differences $\opA - \opAh$ and $\opBh - \opB$
to be proportional to the indentity and thus one would
find the more general matrix-product mechanism of \cite{\HiPeSa}
in the sequential limit.
However, the choice $C = \Ch$ is always possible independent
of the details of the dynamics. This would lead to a linear relation
between the matrices in the sequential limit, but one can
check that the four matrices used in \cite{\HiPeSa} are
linearly independent. Thus, whether the problem including coagulation
and decoagulation admits a matrix-product state also for
parallel dynamics, and if so, with which mechanism, remains
to be investigated.
\mn
The other problem is hopping on a ring with a defect where the
rates are modified. Formally, this case is obtained by replacing the
product $\R \otimes \L$ of the boundary matrices by a
hopping matrix $\widetilde{\T}$. This model has already been
solved by a modified Bethe ansatz for the case of unidirectional
deterministic motion everywhere except at the defect
\cite{\SchuetzBl}. For a certain fixed particle density, the
stationary state is also expressible in the form of a
two-dimensional matrix product \cite{\HiPriv}. One may speculate
that our representation \ref{bulkRep2D} of the bulk algebra may
help to solve the general model with a defect. In particular,
the observation that the differences of the matrices with and
without hat are proportional to the identity is a pure bulk
property and thus one may hope that techniques like those we have used
e.g.\ in Sections 5 and 6 may be useful also for systems with a defect.
\bn
\displayhead{Acknowledgments}
\mn
We thank H.\ Hinrichsen, M.\ Karowski, P.\ Pearce, N.\ Rajewsky,
V.\ Rittenberg and A.\ Zapletal
for useful discussions.  A.H.\ would like to thank the
Deutsche Forschungsgemeinschaft for financial support.
\sectionnumstyle{Alphabetic}
\newsectionnum=0
\vfill
\eject
\appendix{Representation with Jordan form}
\mn
In this appendix we briefly discuss a two-dimensional representation
where $C = \Ch$ has a non-trivial Jordan form. This case occurs
if $C_{2,2} = 1$ and simultaneously $\langle W \state{V} = 0$ because
then the normalization $\astate{W} C^N \state{V}$ of the groundstate
vanishes and the representation with diagonal $C$ may not be used
any more. We have not been able to fully determine when this
happens in \ref{bulkRep2D}, \ref{paramAval} -- \ref{boundVecTwoD},
but one can check that it is the case if $\beta = \alpha$ and
$\gamma = \delta$. Using these values in \ref{RelTwoD}, one then finds that
$$ \left (p{\alpha}^{2}-q{\delta}^{2}\right )\left (1-p\right )\left (1-q
\right )+pq\left (\left (1-p\right )\left (1-\delta\right )^{2}-
\left (1-q\right )\left (1-\alpha\right )^{2}\right )
\ = 0 \, .
\label{RelTwoDjord}$$
One can choose
$$C = \Ch = \pmatrix{1 & 1 \cr 0 & 1 \cr}
\label{jordCan}$$
by suitably fixing the normalization of $C$ and $\Ch$ and the
relative normalization of the two basis vectors in the auxiliary
space.  The further strategy to determine the representation
matrices for $\opA$ und $\opAh$ is now slightly different from
the procedure described in Section 3. First one solves the
boundary conditions \ref{parAlgB} as described below \ref{nonTrivBvec}.
With the result of this computation one can then solve the equations
\ref{parAlgI}. This leads to:
$$\eqalign{
\opA =& {1 \over \abbrevB} \pmatrix{
{q\delta-\delta+p\delta-p}&0 \cr
0&{q\alpha+q\delta-\alpha-q} \cr
} \, , \cr
\opAh =& {1 \over \abbrevB} \pmatrix{
{p\delta-\delta+p\alpha-p} &0 \cr
0&{q\alpha-\alpha+p\alpha-q} \cr
} \, , \cr
}\label{bulkRep2Djord}$$
with
$$\abbrevB =
(\alpha + \delta) (p+q-1)-(p+q) \, .
\label{apprevB}$$
Again, some freedom was used to set $\opAh_{1,2} = 0$ in
\ref{bulkRep2Djord}. Instead of \ref{boundVecTwoD} one now
has to choose the boundary vectors as follows:
$$\eqalign{
\state{V} =&
\pmatrix{
\alpha\,\abbrevB \cr
\left (\alpha+\delta\right )
\left\{ (1-\delta) (1-p) - (1-\alpha)(1-q) \right\} \cr
} \, , \cr
\state{W} =&
\pmatrix{
-\left (\alpha+\delta\right )
\left\{ (1-\delta) (1-p) - (1-\alpha)(1-q) \right\} \cr
\delta\, \abbrevB \cr
} \, .
}\label{boundVecTwoDjord}$$
One can check that $\langle W \state{V}$ is neither zero
nor singular on the manifold given by \ref{RelTwoDjord}.
\mn
A property similar to \ref{matDiffId} also holds here and
has in fact been assumed to simplify the computations.
\mn
We should mention that \ref{RelTwoDjord} is not recovered after
inserting the renormalized rates \ref{renRates} into its
sequential limit. However, it is argued in Section 6 that
the coexistence line has a more general form than just
$\beta = \alpha$, $\gamma = \delta$. Thus, the fact that
\ref{RelTwoDjord} does not reproduce itself under renormalization
is a hint that one should be able to find two-dimensional
representations where $C$ is given by \ref{jordCan} under more
general conditions than \ref{RelTwoDjord}.
\vfill
\eject
\appendix{Symmetric diffusion in the bulk}
\mn
Here we discuss the special case of symmetric diffusion in
the bulk, i.e.\ $p=q$. In this case, which was also considered
in \cite{\KaDoNi}, it is convenient to work with the operator
$\opB$ and the spatial transfer-matrix $C$. Inserting the
choice \ref{chooseDiffId} into \ref{parAlg} leads to
$$\eqalign{
[\opB, C] &= {1 - p \over p} C \, , \cr
\astate{W} \left(\alpha C - (\alpha + \gamma) \opB \right) =
 \astate{W} \, , & \qquad
\left((\beta+\delta) \opB - \delta C \right) \state{V} =
    \left(1 - \beta - \delta\right) \state{V} \, . \cr
}\label{SymmAlg}$$
Using these relations one immediately finds by either
commuting the $\opB$ to the right or the left boundary
$$\eqalign{
\astate{W} C^{x-1} B C^{N-x} \state{V} =&
\left({1 - p \over p} (N-x) + {1 - \beta - \delta \over \beta+\delta}
\right) \astate{W} C^{N-1} \state{V}
+ {\delta \over \beta+\delta} \astate{W} C^N \state{V} \cr
=& \left(-{1 - p \over p} (x-1) - {1 \over \alpha+\gamma}
\right) \astate{W} C^{N-1} \state{V}
 + {\alpha \over \alpha+\gamma} \astate{W} C^N \state{V} \, . \cr
}\label{SymmCompProf}$$
From these two expressions the term $\astate{W} C^{N-1} \state{V}$
can be eliminated yielding the density profile for the odd
sites
$$\langle \tau_x \rangle =
{{\alpha\,\left (\beta+\delta \right )\left (N-1\right )
-\left (\alpha\,\beta - \gamma\,\delta\right ) (x-1) +
{p \over 1-p}\left \{\alpha\,\left (1-\beta\right )
+\delta\,\left (1- \alpha\right )\right\}
\over \left (\alpha+ \gamma\right )\left (\beta+\delta\right )
\left (N-1\right ) +
{p \over 1-p} \left\{\alpha\,\left (1-\beta\right )+\delta
\,\left (1-\alpha\right )+\gamma\,\left (1-\delta\right )+\beta\,
\left (1-\gamma\right )\right\}
}} \, .
\label{SymmResProf}$$
A similar computation with $\opBh = \opB + \id$ yields
the profile for the even sites. The expression differs from
\ref{SymmResProf} only in the last curly bracket in the
numerator which becomes $\left \{\alpha\,\left (1- \delta\right) 
+\delta\,\left (1-\gamma\right )\right\}$.
In the thermodynamic limit $N \to \infty$ this difference
between $\langle \tau_x \rangle$ and $\langle\widehat{\tau}_x\rangle$
disappears and, more strikingly, the densities become independent
of the bulk probability $p$. 
The result after the limit is identical to the one obtained in
\cite{\SchueSti} for the case of sequential updates.
\mn
For finite $N$, the density profile \ref{SymmResProf}
is linear in the spatial variable $x$ (as one can already see from
\ref{SymmCompProf}). For $p=q$ the current is related to
the density profile by $J = p \left(\langle \widehat{\tau}_x \rangle
- \langle \tau_{x+1}\rangle\right)$. Inserting
the result \ref{SymmResProf} one finds that
$$J \approx p {\alpha\,\beta-\gamma\,\delta \over \left(1-p\right)
\left (\alpha+ \gamma\right )\left (\beta+\delta\right )} N^{-1}
\label{SymmCurr}$$
for $N$ large. Physically, this means that the
pumping effect at the ends is not sufficient to drive
a current through an infinite system. The results \ref{SymmResProf}
and \ref{SymmCurr} show that only one phase exists
which corresponds to the coexistence line in Fig.\ 3.
\mn
It is straightforward to apply the method used above for
computing the density profile also to higher correlation functions.
\mn
Although for the case discussed here the groundstate is comparably
simple, this is still interesting because for $p=q$
our local bulk-transfer-matrix $\T$ becomes identical to the $R$-matrix 
$\check{R}$ of the symmetric six-vertex model.
One can parametrize $p=u/(u+\eta)$ where $u$ is the
spectral parameter and $\eta$ a constant. With this
parametrization the Yang-Baxter equation \ref{YBE}
with spectral parameter $u$ is satisfied and can be used
to write down a family of commuting stochastic
row-to-row transfer matrices for periodic boundary conditions
(see e.g.\ Chapter 2.3 of \cite{\Ma}). Thus, the
hopping probability $p$ is essentially the spectral
parameter of an integrable model. One consequence
probably is the disappearance of $p$ from the density
profile \ref{SymmResProf} in the thermodynamic limit.
\mn
For $p \ne q$, the $\T$-matrices \ref{localTransMat} can be identified
with the $\check{R}$-matrix of the general asymmetric six-vertex
model with a suitable spectral parameter $u$:
$$\check{R} =
\pmatrix{
\sin(u+\eta)&0&0&0\cr
0&z^u\,\sin(\eta)&x\,\sin(u)&0\cr
0&x^{-1} \sin(u)&z^{-u}\,\sin(\eta)&0\cr
0&0&0&\sin(u+\eta)\cr
} \, ,
\label{GenRmat}$$
where $\eta$, $x$ and $z$ are some arbitrary constants. One can
show that this also satisfies the Yang-Baxter equation \ref{YBE}.
However, the impact of the boundary vertices in Fig.\ 2 on
the integrability remains to be fully clarified.
\bn
\appendix{Computations on finite chains}
\mn
We show here how to compute correlation functions
efficiently using a representation of the
algebra where \ref{chooseDiffId} holds. Then $\opA$
and $\opB$ satisfy \ref{HamLimAlg} with $g=1$.
A convenient basis for the Fock space is given
by $\opB^x \state{V}$ and one then has to compute
$\opA \opB^x \state{V}$. In order to describe how
this can be done e.g.\ on a computer let
$$\state{x} := \opB^x \state{V} \, , \qquad
\state{x,y} := \opB^x \opA \opB^y \state{V} \, .
\label{FockBasis}$$
The bulk relation in \ref{HamLimAlg} (with $g=1$)
can be used to commute an operator $\opA$ one place
to the right which in terms of the above vectors
yields the relation
$$\state{x,y} = {1 \over q} \left(
p \state{x+1,y-1} - (1-p) \state{x,y-1}
-(1-q) \state{x+y} \right)
\label{CommuteRight}$$
for $y > 0$. As soon as the operator $\opA$ hits the
right boundary one can use the second boundary equation in
\ref{HamLimAlg} with $g=1$ to replace $\opA$ by $\opB$:
$$\state{x,0} = {1 \over \delta} \left(
     \beta \state{x+1} - (1 - (\beta + \delta)) \state{x}
    \right) \, .
\label{CommuteRiBound}$$
By definition, the algebra acts on the states
\ref{FockBasis} as
$$\opA \state{x} = \state{0,x} \, , \qquad
\opB \state{x} = \state{x+1} \, .
\label{FockAction}$$
The rules \ref{CommuteRight} -- \ref{FockAction} are sufficient
to express any state $\opA \opB \opA \opA \ldots \state{V}$ in
terms of the $\state{x}$ and one does not need the original algebra
anymore. In order to be able to compute words (i.e.\ scalar products
with $\astate{W}$) a final constant $s_x := \langle W \state{x}$
is needed. It can be computed by creating an $\opA$
at the left boundary using the first boundary relation in
\ref{HamLimAlg} (with $g=1$). This leads to
$$s_{x+1} = {1 \over \gamma} \left(\alpha \langle W \state{0,x}
            - s_x \right)
\label{RecurrSx}$$
which amounts to a recurrence relation for the $s_x$ after
moving the $\opA$ to the right boundary using the previous
relations.  Setting $s_0 = 1$, one is now able to compute the value
of any word $\astate{W} \opA \opB \opA \opA \ldots \state{V}$
exclusively from the rules \ref{CommuteRight} -- \ref{RecurrSx}.
\mn
It is possible to solve these recurrence relations in closed
form following the lines of \cite{\Sandow}. However, also
formulae of the type as in \cite{\Sandow} are best evaluated
numerically using a recursive procedure. The recipe presented
above is sufficient to compute correlation functions
numerically on finite chains where a length $N=100$ is no
major problem. However,   
in order not to do a computation twice, one should store
the expansion of the vectors $\state{x,y}$ in terms of the
basis vectors $\state{x}$. One also needs to be careful
with the numerical range because some of the numbers
grow exponentially with $N$, e.g.\
$\astate{W} C^N \state{V} \approx J^{-N}$ and
$\abs{J^{-1}} \ge 1$ can become quite large
\footnote{${}^{3})$}{An implementation
in C which takes care of all these details is available
on the WWW under URL
\hbox{
\href{http://www.physik.fu-berlin.de/~ag-peschel/software/mp.html}
{http://www.physik.fu-berlin.de/\~{}ag-peschel/software/mp.html}}\ .
This program does not only compute he current, the
density profile and the two-point function on a finite chain,
but also implements our results in Table 1 as well as
the correlation length \ref{resCorrLen} and thus provides a simple
way for checking their validity numerically.
}.
\mn
Fig.\ 5 shows a density profile obtained in this manner on
a finite lattice with $N=200$. The parameters yield
$\kappah_{+}(\alpha,\gamma) = 0.912$ and
$\kappah_{+}(\beta,\delta) = 0.801$ which to corresponds
to a point at the top right corner of phase III in
Fig.\ 3. The distinction between the two sublattices is
clearly visible, so is the influence of the boundaries.
As a byproduct in this computation one finds the current
with these parameters for $N = 200$ using \ref{currentFormula}:
$J = 0.2690$. This is to be compared to the result
$J = 0.2679$ obtained from \ref{maxCurrent} in the
thermodynamic limit.
Computations like this one provide room for more
detailed investigations.
\mn
\centerline{\psfig{figure=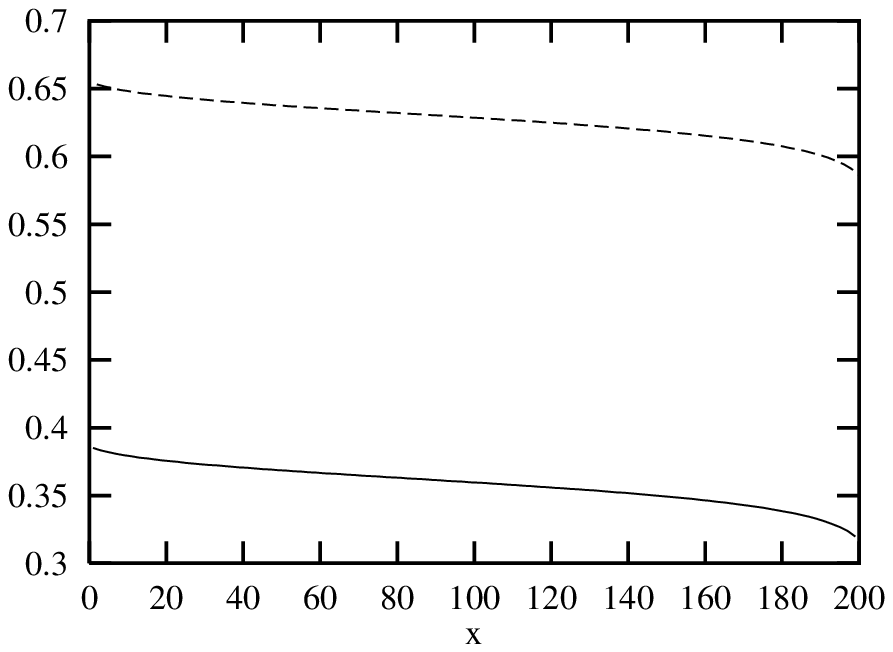}}
\sn
{\par\noindent\figindents
{\bf Fig.\ 5:} A density profile in the maximal
current phase with $N = 200$ sites and
$p = 0.75$, $q = 0.25$, $\alpha = 0.5$, $\beta = 0.6$,
$\gamma = 0.1$ and $\delta = 0.2$. The full line
shows the density $\langle\tau_{2x+1}\rangle$
and the dashed line the density $\langle\widehat{\tau}_{2x}\rangle$.
\par\noindent}
\vfill
\eject
\displayhead{References}
\mn
\bibitem{\DDM} B.\ Derrida, E.\ Domany, D.\ Mukamel, {\it An Exact Solution of
              a One Dimensional Asymmetric Exclusion Model with Open
              Boundaries}, J.\ Stat.\ Phys.\ {\bf 69} (1992) 667-687
\bibitem{\DoSch} E.\ Domany, G.M.\ Sch\"utz, {\it Phase Transitions in an
              Exactly Soluble One-Dimensional Exclusion Process}, J.\ Stat.\
              Phys.\ {\bf 72} (1993) 277-296
\bibitem{\DEHP} B.\ Derrida, M.R.\ Evans, V.\ Hakim, V.\ Pasquier, {\it Exact
              Solution of a 1D Asymmetric Exclusion Model Using a Matrix
              Formulation}, J.\ Phys.\ A: Math.\ Gen.\ {\bf 26} (1993)
              1493-1518
\bibitem{\DeEv} B.\ Derrida, M.R.\ Evans, {\it Exact Steady State Properties
              of the One Dimensional Asymmetric Exclusion Model}, In:
              G.\ Grimmett (ed.), {\it Probability and Phase Transition},
              Kluwer Academic Publishers (1994) 1-16           
\bibitem{\HaNa} V.\ Hakim, J.P.\ Nadal, {\it Exact Results for 2D Directed
              Animals on a Strip of Finite Width}, J.\ Phys.\ A: Math.\ Gen.\
              {\bf 16} (1983) L213-L218
\bibitem{\KSZa} A.\ Kl\"umper, A.\ Schadschneider, J.\ Zittartz, {\it
              Equivalence and Solution of Anisotropic Spin-1 Models and
              Generalized $t-J$ Fermion Models in One Dimension}, J.\ Phys.\
              A: Math.\ Gen.\ {\bf 24} (1991) L955-L959
\bibitem{\KSZb} A.\ Kl\"umper, A.\ Schadschneider, J.\ Zittartz, {\it Matrix
              Product Ground States for One-Dimensional Spin-1 Quantum
              Antiferromagnets}, Europhys.\ Lett.\ {\bf 24} (1993) 293-297
\bibitem{\HiPeSa} H.\ Hinrichsen, I.\ Peschel, S.\ Sandow, {\it On Matrix Product
              Ground States for Reaction-Diffusion Models},
              J.\ Phys.\ A: Math.\ Gen.\ {\bf 29} (1996) 2643-2649
\bibitem{\Hinrichs} H.\ Hinrichsen, {\it Matrix Product Ground States for Exclusion
              Processes with Parallel Dynamics}, preprint cond-mat/9512172
\bibitem{\PeRy} I.\ Peschel, F.\ Rys, {\it New Solvable Cases for the
              Eight-Vertex Model}, Phys.\ Lett.\ {\bf A91} (1982) 187-189
\bibitem{\Rujan} P.\ Ruj\'an, {\it Order and Disorder Lines in Systems with
              Competing Interactions: II.\ The IRF Model}, J.\ Stat.\ Phys.\
              {\bf 29} (1982) 247-262
\bibitem{\Baxter} R.J.\ Baxter, {\it Disorder Points of the IRF and Checkerboard
              Potts Models}, J.\ Phys.\ A: Math.\ Gen.\ {\bf 17} (1984)
              L911-L917
\bibitem{\BaLe} M.T.\ Batchelor, J.M.J.\ van Leeuwen, {\it Disorder Solutions
              of Lattice Spin Models}, Physica {\bf A154} (1989) 365-383
\bibitem{\RSS} N.\ Rajewsky, A.\ Schadschneider, M.\ Schreckenberg, {\it The
              Asymmetric Exclusion Model with Sequential Update}, preprint
              cond-mat/9603172, to appear in J.\ Phys.\ A: Math.\ Gen.\
\bibitem{\KaDoNi} E.\ Domany, D.\ Kandel, B.\ Nienhuis, {\it A Six-Vertex Model
              as a Diffusion Problem: Derivation of Correlation Functions},
              J.\ Phys.\ A: Math.\ Gen.\ {\bf 23} (1990) L755-L762
\bibitem{\Schuetz} G.M.\ Sch\"utz, {\it Time-Dependent Correlation Functions in a
              One-Dimensional Asymmetric Exclusion Process}, Phys.\ Rev.\
              {\bf E47} (1993) 4265-4277
\bibitem{\OwBa} R.J.\ Baxter, A.L.\ Owczarek, {\it Surface Free Energy of the
              Critical Six-Vertex Model with Free Boundaries}, J.\ Phys.\ A:
              Math.\ Gen.\ {\bf 22} (1989) 1141-1165
\bibitem{\Sandow} S.\ Sandow, {\it Partially Asymmetric Exclusion Process with
              Open Boundaries}, Phys.\ Rev.\ {\bf E50} (1994) 2660-2667
\bibitem{\EssRi} F.H.L.\ E{\ss}ler, V.\ Rittenberg, {\it Representations of the
              Quadratic Algebra and Partially Asymmetric Diffusion with Open
              Boundaries}, preprint cond-mat/9506131, BONN-TH-95-13,
              version of April 1996
\bibitem{\ADHR} F.C.\ Alcaraz, M.\ Droz, M.\ Henkel, V.\ Rittenberg, {\it
              Reaction-Diffusion Processes, Critical Dynamics, and Quantum
              Chains}, Ann.\ Phys.\ {\bf 230} (1994) 250-302
\bibitem{\KoelnB} N.\ Rajewsky, L.\ Santen, A.\ Schadschneider, M.\ Schreckenberg,
              in preparation
\bibitem{\Ma} Z.-Q.\ Ma, {\it Yang-Baxter Equation and Quantum Enveloping
              Algebras}, World Scientific, Singapore (1993)
\bibitem{\GoVe} A.\ Gonz\'alez-Ruiz, H.J.\ de Vega, {\it Boundary $K$-Matrices
              for the XYZ, XXZ and XXX Spin Chains}, J.\ Phys.\ A: Math.\
              Gen.\ {\bf 27} (1994) 6129-6138
\bibitem{\InKo} T.\ Inami, H.\ Konno, {\it Integrable XYZ Spin Chain with
              Boundary}, J.\ Phys.\ A: Math.\ Gen.\ {\bf 27} (1994) L913-L918
\bibitem{\BaYu} M.T.\ Batchelor, C.M.\ Yung, {\it Integrable Vertex and Loop
              Models on the Square Lattice with Open Boundaries via
              Reflection Matrices}, Nucl.\ Phys.\ {\bf B435} (1995) 430-462
\bibitem{\Giac} H.J.\ Giacomini, {\it Disorder Solutions and the Star-Triangle
              Relation}, J.\ Phys.\ A: Math.\ Gen.\ {\bf 19} (1986) L537-L541
\bibitem{\HiKrPe} H.\ Hinrichsen, K.\ Krebs, I.\ Peschel, {\it Solution of a
              One-Dimensional Diffusion-Reaction Model with Spatial
              Asymmetry}, Z.\ Phys.\ {\bf B100} (1996) 105-114
\bibitem{\SchuetzBl} G.M.\ Sch\"utz, {\it Generalized Bethe Ansatz Solution of a
              One-Dimensional Asymmetric Exclusion Process on a Ring with a
              Blockage}, J.\ Stat.\ Phys.\ {\bf 71} (1993) 471-505
\bibitem{\HiPriv} H.\ Hinrichsen, private communication
\bibitem{\SchueSti} G.M.\ Sch\"utz, R.B.\ Stinchcombe, {\it Operator Algebra for
              Stochastic Dynamics and the Heisenberg Chain}, Europhys.\ Lett.\
              {\bf 29} (1995) 663-667
\vfill
\end